\documentclass[reprint, amsmath,amssymb, aps]{revtex4-1}

\usepackage{times}
\usepackage{graphicx}
\usepackage{subfig}
\usepackage{epstopdf}
\usepackage{psfrag}
\usepackage{ae}
\usepackage{amsmath,amssymb}
\usepackage{float}
\usepackage[usenames]{color} 
\usepackage[dvipsnames]{xcolor}
\usepackage[normalem]{ulem}
\begin{document}
	
	\title{Testing theories of the glass transition with the same liquid, but many kinetic rules}
	
	\author{Cristina Gavazzoni} 
	\email{crisgava@gmail.com}
	\affiliation{Instituto de F\'isica, Universidade Federal
		do Rio Grande do Sul, Caixa Postal 15051, CEP 91501-970, 
		Porto Alegre, Rio Grande do Sul, Brazil}
	
	\author{Carolina Brito}
	\email{carolina.brito@ufrgs.br}
	\affiliation{Instituto de F\'isica, Universidade Federal
		do Rio Grande do Sul, Caixa Postal 15051, CEP 91501-970, 
		Porto Alegre, Rio Grande do Sul, Brazil}
	
	\author{Matthieu Wyart} 
	\email{matthieu.wyart@epfl.ch}
	\affiliation{Institute of  Physics, Ecole Polytechnique 
		Federale de Lausanne, 729 BSP UNIL,
		1015, Lausanne, Switzerland}
	
	\date{\today}
	
	\begin{abstract}
		We study the glass transition by exploring a broad  class of kinetic rules that can   significantly modify the normal dynamics of super-cooled liquids, while maintaining thermal equilibrium. Beyond the usual dynamics of liquids, this class  includes  dynamics in which a fraction $(1-f_R)$ of the particles can perform pairwise exchange or `swap moves', while a fraction $f_P$  of the  particles can only move along restricted directions.  We find  that (i)  the location of the glass transition varies greatly but smoothly as $f_P$ and $f_R$ change and (ii) it is governed by a linear combination of $f_P$ and $f_R$. (iii) Dynamical heterogeneities (DH) are not governed by the static structure of the material,  their magnitude correlates instead with the relaxation time. (iv) We show that a recent theory for temporal growth of DH  based on thermal avalanches holds quantitatively throughout the $(f_R,f_P)$ diagram. 
		These observations are negative items for some existing theories of the glass transition, particularly those reliant on growing thermodynamic order or locally favored structure, and open new avenues to test other approaches, as we illustrate.
	\end{abstract}
	
	\maketitle
	
	Understanding why liquids glass formers cease to flow near their glass transition $T_g$ remains a challenge. At that  point, the relaxation time $\tau_\alpha$ beyond which stress relaxes is of order of minutes, which is fifteen decades larger than at high temperatures. From $\tau_\alpha$, the activation energy  $E_a$ can be defined as  $\tau_\alpha=t_0 \exp(E_a/T)$, where  $t_0$ is a microscopic time scale and $T$ is the temperature (in the units of the Boltzmann constant). In liquids called fragile, $E_a$ can increase five-fold or more under cooling \cite{Anderson95,ediger1996supercooled,Debenedetti01,berthier2011theoretical}. As the dynamics slows down, it also becomes more and more heterogeneous, corresponding to a growing  length scale $\xi$ ~\cite{kob1997dynamical,yamamoto1998dynamics,dalle2007spatial,karmakar2014growing}. Contrasting theories seek to explain these two facts. In the first class of views,  including Adam-Gibbs \cite{adam1965temperature} and Random First Order Theory  (RFOT) \cite{kirkpatrick1989scaling,lubchenko2007theory,Biroli12},  the increase of activation energy stems from the emergence of  some order on a growing length $\xi$, that must be destroyed by cooperative motion on that scale to relax the material. $E_a$  can then be expressed in terms of purely thermodynamic quantities, independently of the kinetic rules governing the dynamics.  Some real space approaches associate such a growing order to locally favored structures \cite{patrick2008direct,kawasaki2007correlation}. A second viewpoint  seeks to capture the mechanism of dynamical facilitation,  whereby the relaxation of a given region speeds up the relaxation of regions nearby.  Kinetically constrained models \cite{ritort2003glassy,berthier2011dynamical}, such as the East model, capture  this effect and suggest a scenario  \cite{garrahan2002geometrical,garrahan2007dynamical,hedges2009dynamic} in which thermodynamics plays almost no role, but dynamics is heterogeneous and the growth of activation energy stems from non-local rearrangements taking place over $\xi$.  At odds with these two views, a third approach that includes free volume \cite{turnbull1961free} or elastic \cite{dyre2006colloquium,JeppeEdan,lerbinger2022relevance,massimo} models, assumes that the activation energy is not controlled by a growing length scale. Instead, it is governed by the energy cost of elementary rearrangements of a few particles jumping over a barrier. The elastic coupling between rearrangements \cite{Lemaitre14,chowdhury2016long,tong2020emergent,wu2015anisotropic,maier2017emergence,steffen2022molecular,flenner2015long,klochko2022theory,chacko2021elastoplasticity} leads to a correlated dynamics ~\cite{ozawa2023elasticity} that can be described in terms of avalanches of activated events \cite{tahaei2023scaling}.
	
	\begin{figure*}
		\includegraphics[width=0.8\textwidth]{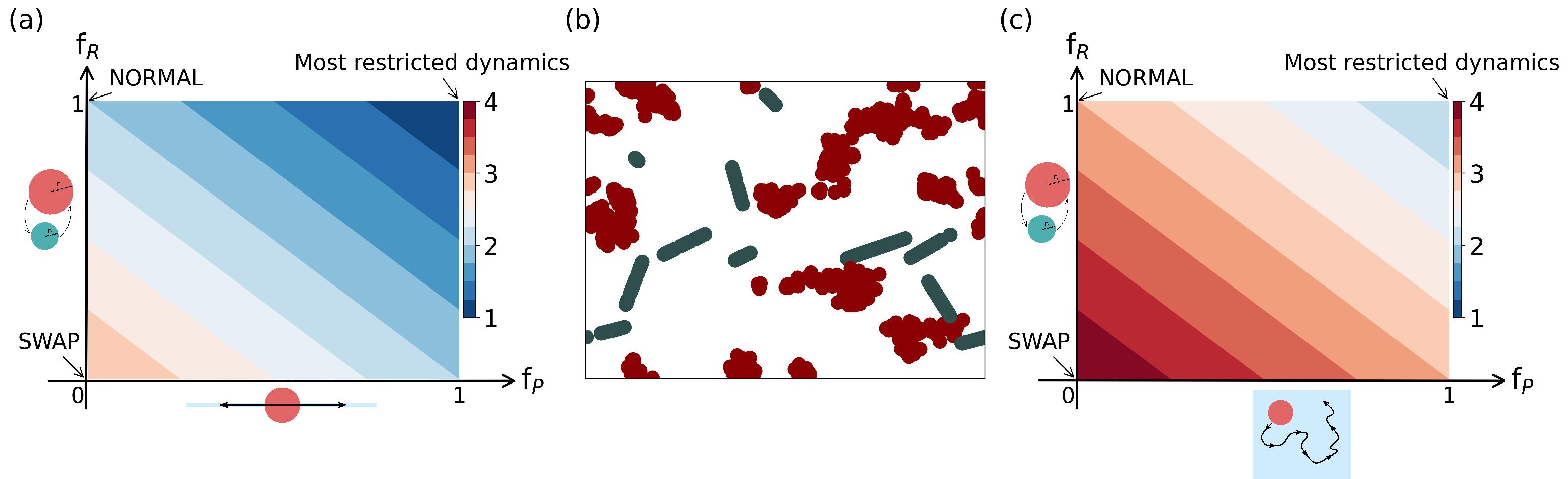}
		\caption{\label{planos}%
			Diagram indicating the number of degrees of freedom  for spatial dimension $d=2$ (a) and $d=3$ (c) as a function of the fraction $f_R$ of particles that cannot swap  and the fraction  $f_P$ of particles whose translation motion is restricted.   Panel (b) shows the particles positions at various time points within some time interval for $d=2$. Particles that can move freely in all direction are shown in red, and particles that are restricted to linear motion appear in blue. For $d=3$, this constraint is more gentle, as particles can still move on planes as sketched in (c).}
	\end{figure*}
	
	Molecular dynamics simulations of models of super-cooled liquids have been extremely informative to characterize the glass transition \cite{kob1999computer}, yet  distinct views on this phenomenon have been hard to definitely contrast \cite{Biroli12}. Our present goal is to show that for some popular models of liquids, a very broad class of kinetic rules  can be considered, which can continuously (and very significantly) speed up or slow-down the normal dynamics, while preserving thermal equilibrium. Although these rules would be hard to implement in actual experiments, they are equivalent to  dynamics with purely local rules, and as such theories of the glass transition should  apply to them. This approach thus opens an avenue to test more stringently theories of glassy dynamics. Specifically, our work builds on  `swap' Monte Carlo algorithms. In these algorithms,   pairs of particles can exchange positions, in addition to their usual translation moves  \cite{Glandt84,Grigera01fast,fernandez2007phase,gutierrez2015static,ninarello2017models}. For continuously polydisperse systems, these algorithms can speed up the dynamics by 15 orders of magnitude or more \cite{ninarello2017models},  and can change the  glass transition temperature $T_g$ by up to a factor of two.  It allows one  to explore glasses with a stability similar to that reached in experiments. 
	
	Our central result is to introduce a family of kinetic rules, where a fraction $f_R$ of the particles cannot swap, and a fraction $f_P$ of the particles can only move along  randomly-chosen hyperplanes. We provide systematic measurements of the dynamics in the $(f_R, f_P)$ diagram in two and three dimensions, that includes the normal dynamics $(1,0)$ as well as swap $(0,0)$. Overall, our observations are negative items for theories based on a growing thermodynamic order.  We discuss how devoted studies could be used in these models to test alternative views of the glass transition.
	
	{\bf Changing continuously the kinetic rules of liquids:} 
	Swap moves lead to a considerable speed up of the dynamics \cite{Glandt84,gutierrez2015static,ninarello2017models}. Importantly, despite its apparent non-local character, swap dynamics can be conceived as a purely local dynamics. Following \cite{Brito18,Glandt84,hagh2021transient}, swap is equivalent to considering identical particles endowed with an additional `breathing' degree of freedom, allowing them to change their size according to some chemical potential $\mu(R)$. Indeed, letting  pairs of particles exchange is equivalent, in the thermodynamic limit, to letting individual particles exchange with a bath of particles of all possible sizes $R$.  $\mu(R)$ is then chosen to obtain the desired polydispersity, which is continuous for continuously poly-disperse particles \cite{Brito18,Glandt84}.
	Adding such a degree of freedom per particle  dramatically softens the energy landscape \cite{szamel2019theory,ikeda2019mean,Brito18,Wyart17} while preserving thermodynamic and structural properties. It also affects the dynamics: following  the center of the particles, and considering that a swap move corresponds to a change of the  size of two particles but not of their position, leads to the following observation.  Dynamical correlations grow under cooling as for the normal dynamics, but the correlation length starts growing  at a much smaller temperature \cite{ninarello2017models}. 
	To study more systematically such effects and their consequences, 
	following \cite{gopinath2022diffusion} we vary the parameter $f_R\in [0,1]$, characterizing the number of particles that cannot swap. Unlike \cite{gopinath2022diffusion}, we do not consider that particles that swap positions  perform a jump, as it leads to very different dynamics.
	
	Following this logic, we propose to add even more kinetic constraints by restricting the motion of a fraction $f_P\in [0,1]$ of the  particles, as illustrated in Fig.\ref{planos}. Each such  particle is forbidden  to move along one random direction associated to it, and for an infinite system they would be each confined  along a random hyperplane. Overall, the number of degrees of freedom for a system of $N$ particles is then $N(d-f_P+1-f_R)$. Note that for the periodic boundary condition we consider below, this dynamics is ergodic as such hyperplanes visit the neighborhood of any point with probability one, if their orientation is randomly chosen. For dynamics that satisfy detailed balance like ours, it ensures that thermal equilibrium will eventually be reached. Thus structural and thermodynamic properties are only governed by $\phi$, independently of the  choices of kinetic rules embodied in $(f_R,f_P)$. Note that other procedures were proposed to reduce  the number of degrees of freedom, such as  pinning particles starting from an equilibrated system as proposed  e.g. in \cite{cammarota2012ideal}.  Yet in that case ergodicity is obviously broken once pinned particles are chosen, and the dynamical properties of the system are not translation-invariant anymore.
		
	In the present work, we specifically perform Monte Carlo simulations of systems with $N$ continuously polydisperse hard spheres particles of packing fraction $\phi$ in a regular box of linear size $L$, with periodic boundary conditions. For hard particles, instead of temperature, $\phi$ is the good controlled parameter. The choice of polydispersity {together with other numerical details are}  shown in the Appendix \ref{det}. Figure \ref{planos}-(a,c) shows a schematic diagram of the different dynamics we explore in $d=2$ and $d=3$ respectively, and indicate in color the associated number of degrees of freedom.  Figure\ref{planos}-(b) illustrates an example of the particles trajectories at a short time for $d=2$, revealing which particles are restricted to linear motion, and which ones are not.

		\begin{figure*}
			\includegraphics[width=0.9\textwidth]{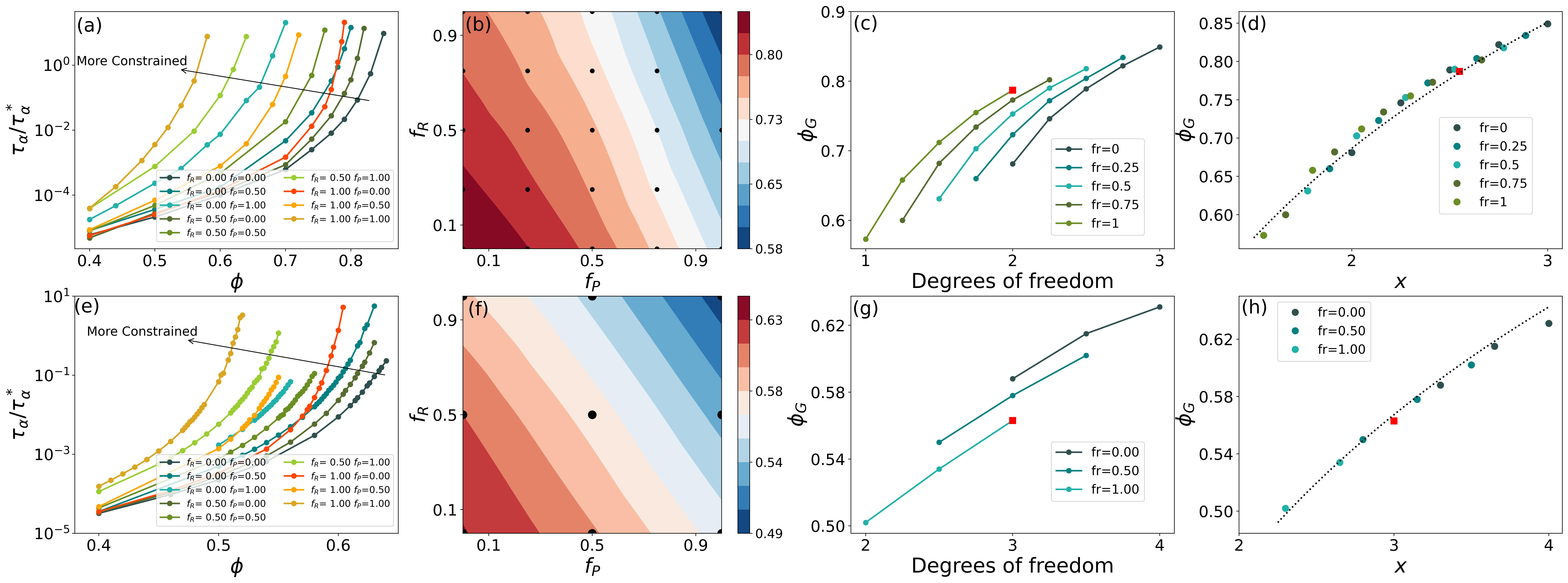} 
			\caption{$\tau_{\alpha}/\tau_{\alpha}^*$ as function of $\phi$ for different values of $f_R$ and $f_P$ for $d=2$ (a) and $d=3$ (e). From these curves, $\phi_G$ as a function of $f_R$ and $f_P$ is extracted, as indicated in color for  $d=2$ (b) and  $d=3$ (f). The discs mark  the simulation data, from which extrapolations are made.  
			(c,g) show $\phi_G$ as a function of the number of degrees of freedom per particle $d+1-f_R-f_P$ and (d,h) show that $\phi_G$ can be collapsed as a function of an effective number of constraints $x=f_P+C(d)f_R$, with $C(2)=0.45$, and $C(3)= 0.70$. The normal dynamics is marked as a red square in these plots.
			}
			\label{fp_fr}
		\end{figure*}

	{\bf Dependence of the glass transition packing fraction $\phi_G$ with kinetic rules:}
	For $d=3$, our choice of polydispersity and Monte-Carlo algorithm follows closely previous works \cite{berthier2016equilibrium, ninarello2017models}, where it was shown that swap can speed up the dynamics by 15 decades or more.  Here we observe a giant speed up of swap in our $d=2$ system as well, as documented in Appendix \ref{appContSwap}. We consider the dynamics on the full phase diagram $(f_R,f_P)$. Fig.\ref{fp_fr}-(a,e)  show $\tau_{\alpha}$ {\it vs} $\phi$ for such dynamics as the parameters $(f_R,f_P)$ are varied, both for $d=2$ and $d=3$.  Fig.\ref{fp_fr}-(b,f) represents the corresponding value of $\phi_G$ in color, as extrapolated from the different measurements made as indicated by circles. The most remarkable results are that (i) $\phi_G$ varies very significantly, and continuously throughout the phase diagram. In particular, there is no evidence for a region surrounding the normal dynamics where  kinetic constraints would not matter and $\phi_G$ would plateau. The normal dynamics can be made continuously faster or slower. (ii) Observing these two  diagrams, it is apparent that most of the variation of $\phi_G$ is captured by a linear combination of $f_R$ and $f_P$.This hypothesis is tested in Fig.\ref{fp_fr}-(d,h), where it is shown that $\phi_G(f_R,f_P)=\phi_G(x)$, where $x$ can be thought of as an effective number of  constraints $x=f_P+C(d) f_R$, where  the coefficient  $C(d)\leqslant 1$  characterize the relative effect of breathing degrees of freedom {\it v.s.} translational ones. (iii) qualitative observations are independent of the spatial dimension $d$.
		
		\begin{figure}
			\includegraphics[width=0.45\textwidth]{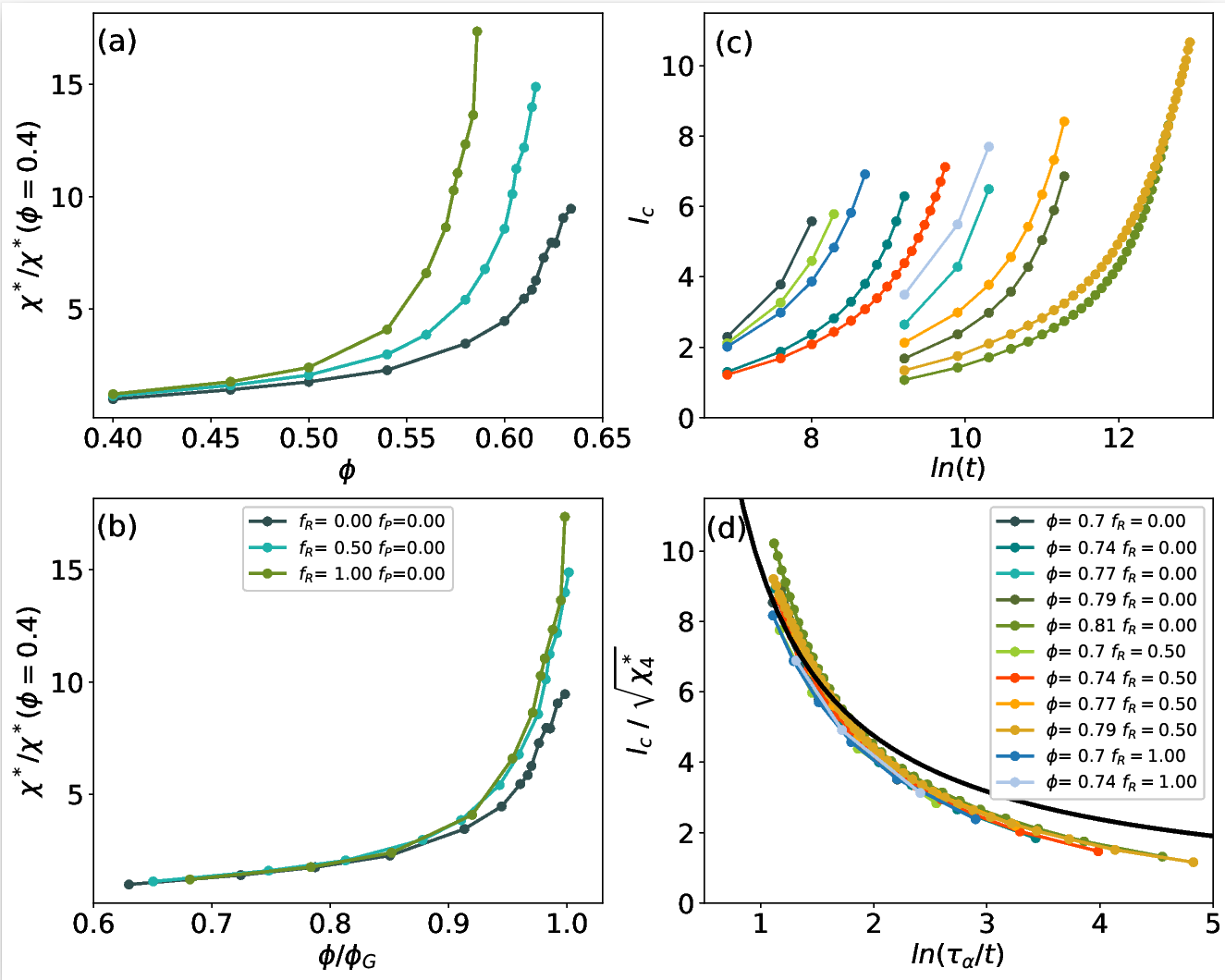}
			\caption{
				{\bf (a)} Maximum $\chi_4^*$ of the dynamic susceptibility $\chi_4(t)$ for $d=3$ as a function of $\phi$ as $f_R$ as  is varied indicated in legend, and $f_P=0$.  {\bf (b)} $\chi_4^*/\chi_4{^*}(\phi=0.4)$ as a function of $\phi/\phi_G$. Comparing (a) and (b) reveals that  $\phi/\phi_G$ is a much better predictor of DH than $\phi$. {\bf (c)}  Coarsening length $l_c$ for $d=2$ {\it v.s. } $\ln(t)$ as $\phi$  is varied, for different values of $f_R$ indicated in legend {\bf (d)} for $f_P=0$ and $t \le \tau_{\alpha}/3$.   {\bf (d)}  $l_c$ collapses when axis are rescaled as predicted by Eq.\ref{eq:prediction_for_ell}. The full black line is the theoretical prediction $l_c = a \sqrt{\chi_4^*} [\ln(\tau_{\alpha}/t)]^{-1}$ with  $a=9.5$. }
			\label{snap}
		\end{figure}

	{\bf Dynamical heterogeneities are not governed by structural properties:}
	Various theoretical approaches propose that dynamical heterogeneities are controlled by equilibrated structural properties, such as  the extension of locally-favored structure \cite{patrick2008direct,kawasaki2007correlation} or  a point to set length \cite{lubchenko2007theory,Biroli12} entering RFOT (although additional effects such as facilitation can be added to this theory to increase further dynamical correlations, see e.g. \cite{biroli2023rfot} for a recent discussion). In our set-up, these properties depend only on $\phi$, and not on the values of parameters $f_R,f_P$.  Here instead we find that for the relaxation times we can probe and the system we consider,  the dynamical susceptibility $\chi_4$ characterizing the magnitude of dynamical heterogeneities are not governed by $\phi$ only, but instead depend strongly on $f_R,f_P$ as illustrated in Fig.\ref{snap}(a). Much less variation in $\chi_4$ is found when plotted as a function of $\phi/\phi_G$ as shown in Fig.\ref{snap}(b).
		
	These results indicate that heterogeneities are not controlled by a length scale that would directly appear in the structure. Indeed, dynamical correlations  are small for the swap dynamics at significant packing fraction for which the normal dynamics is already very correlated \cite{ninarello2017models}. Thus, although locally favored structure may affect dynamical heterogeneities in specific systems (such as two-dimensional systems of discs that can display hexatic order \cite{kawasaki2007correlation}), it does not appear to be the case for continuously polydisperse particles, at least in the range of time scale that can be probed numerically.

	{\bf The temporal evolution of DH follows that of thermal avalanches:}
	Starting from a reference equilibrated configuration, coloring rearranging regions reveals a  coarsening phenomenon \cite{chandler2010dynamics,scalliet2022thirty} and a growing length scale $l_c(t)$, as shown in Fig.\ref{snap}.  Quantitatively we define $l_c(t)=\langle l^3 \rangle_t/\langle l^2 \rangle_t$ where the average $\langle  \rangle_t$ is made on all clusters of connected relaxed particles at time $t$, and $l$ is the square root of the number of particles involved in a cluster. In \cite{tahaei2023scaling}, a description of this coarsening based on facilitation via thermal avalanches of local activated rearrangements was proposed. It was refined  and also argued to apply to  creep flows in elastic manifolds  in \cite{de2024dynamical}, and implies for $t\ll\tau_\alpha$ and $l_c$ much smaller than the system size:
	
			\begin{equation}
				l_c \sim \xi \left[\ln(\tau_\alpha/t)\right]^{-1/(\tilde \sigma \tilde d_f)}
				\label{eq:prediction_for_ell}
			\end{equation}
	
	where $\xi$ is the correlation length, $\tilde \sigma$ some exponent and $\tilde d_f$ is the fractal dimension of the clusters. In this view, $\xi^{\tilde d_f}\sim \chi_4^*$ which characterizes the typical cluster volume.  For $d=2$, $\tilde d_f\approx 2$ and $\tilde \sigma \approx 1/2$ \cite{tahaei2023scaling}, leading to $l_c \sim \sqrt{\chi_4^*} \left[\ln(\tau_\alpha/t)\right]^{-1}$. As shown in Fig.\ref{snap}(d), this prediction works remarkably well, independently of  $f_R$. It supports that thermal avalanches cause DH independently of the kinetic rules chosen.

		{\bf Discussion:}
		Swap Monte Carlo algorithms can be restricted to local moves with no significant effects on the dynamics \cite{ninarello2017models}. More fundamentally, they are equivalent to a purely local kinetic rule where particles can adapt their radii  \cite{Glandt84,Brito18}.  For theories based on a growing order on some  length $\xi_{coop}$ such as RFOT, or approaches based on locally  favored structures, barriers are cooperative and can be expressed in terms of thermodynamic quantities alone.  They should be present for any local kinetic rules  \cite{Wyart17}, including those studied here.  Thus in these approaches, the core mechanism slowing down the dynamics near the glass transition should not depend on the choice of $(f_R,f_P)$. Authors in \cite{Berthier19} acknowledged that swap and normal dynamics should asymptotically relax at the same pace according to RFOT,  but countered that pre-asymptotic corrections (not currently described within this theory) may cause the observed difference.

		The main difficulty with this view is that the dynamics is very different as  the parameters $(f_R,f_P)$ change: as shown in Fig.\ref{cont_swap}, these dynamics do not become equivalent even after a slowing down of 15 decades accessible experimentally. To have predictive power,   RFOT or theories based on thermodynamic quantities should  specify a value for the parameters $(f_R,f_P)$ for which they  apply. However, currently they don't. 
		Our observation  that $\phi_G$ continuously  depends on  $(f_R,f_P)$, and does not plateau to some constant value around the normal dynamics $(1,0)$, shows that the normal dynamics is just one among many other dynamics. This point underlines the lack of predictive power of RFOT or related theories - at least for the systems of continuously polydisperse particles studied here. 
		
		By contrast, for theories based on kinetic constraints or on local barriers (known to depend on kinetic rules \cite{Brito18}), the fact that $\phi_G$ should continuously  vary with the amount of constraints is evident. The normal dynamics is slower simply because it is a kinetically constrained version of swap dynamics. 

		{\bf Conclusion:}
		RFOT is a mean-field theory of the glass transition, which has shown undeniable successes. It is exact in infinite dimension \cite{Charbonneau14a,Maimbourg16},  correctly captures aspects of the thermodynamics of super-cooled liquids \cite{biroli2008thermodynamic}, and presents a dynamical transition \cite{kurchanlaloux,Biroli06,lubchenko2007theory} akin to mode coupling theory, that describes some aspects of liquid dynamics at intermediate temperatures \cite{gotze2008complex}. Yet our results support that its description of activation near the glass transition does not apply for the continuously polydisperse systems studied here. Although our conclusions are restricted to these specific systems where swap is so performant, these models are known to capture the key facts associated with the glass transition \cite{scalliet2022thirty}.
		
		
		The model we introduced, with its very large variation of dynamics with different kinetic rules, offers new opportunities to test  
		different theories of the glass transition, extending previous observations that were only considered for a single kinetic rule. As an illustration, we showed that the time evolution of dynamical heterogeneities follow a description of facilitation based on thermal avalanches of local rearrangements \cite{tahaei2023scaling} independently of the kinetic rules chosen, emphasizing the robustness of this theory. Other debated issues that this model would help resolve include  what observable predicts best the regions which will flow first, the so-called dynamical propensity, for which many candidates were proposed  \cite{widmer2004reproducible,lerbinger2022relevance,alkemade2022comparing,schoenholz2017relationship}. 
		Likewise, numerical tests put forward to test theories of the slowing down of the dynamics (i.e. how the activation energy $E_a$ depends on temperature or packing fraction)
		can now be made much more stringent. The list includes kinetically constrained or lattice gas models \cite{toninelli2006jamming}, that can be tested  by analyzing irreversible events  \cite{keys2011excitations}. Furthermore, the notion that glassy dynamics correspond to local rearrangements  was supported by the measurement of the density of state of local barriers \cite{massimo}; and these barriers were argued to be governed alternatively by global \cite{dyre2006colloquium} or  local \cite{JeppeEdan} elasticity,  or by the  amplitude of vibrational motion \cite{puosi2012communication}. Varying $(f_R,f_P)$ in these measurements will indicate which viewpoint is most likely correct. Overall, we have added an axis to the glass transition problem by varying continuously kinetic rules, affecting strongly observations and giving a new handle to  decide which theory of the glass transition actually applies. 
		
		\begin{acknowledgments}
			We thank the Simons collaboration for discussions, in
			particular L. Berthier, G. Biroli, M. Ozawa and C. Scalliet. We thank J.P. Bouchaud, T. deGeus, W. Ji, M.
			Muller, M. Pica Ciamarra, M. Popovic, A. Tahei and A.
			Rosso for discussions, and J. Kurchan for exchanges at
			the beginning of this project. MW acknowledges sup-
			port from the Simons Foundation Grant (No. 454953
			Matthieu Wyart) and from the SNSF under Grant No.
			200021-165509. C.B. and C.G. thank the Brazilian agency CAPES and CNPq for the financial support.. 
			
		\end{acknowledgments}
		
		
		\bibliographystyle{apsrev4-1}
		\bibliography{refs}

\begin{thebibliography}{67}%
\makeatletter
\providecommand \@ifxundefined [1]{%
 \@ifx{#1\undefined}
}%
\providecommand \@ifnum [1]{%
 \ifnum #1\expandafter \@firstoftwo
 \else \expandafter \@secondoftwo
 \fi
}%
\providecommand \@ifx [1]{%
 \ifx #1\expandafter \@firstoftwo
 \else \expandafter \@secondoftwo
 \fi
}%
\providecommand \natexlab [1]{#1}%
\providecommand \enquote  [1]{``#1''}%
\providecommand \bibnamefont  [1]{#1}%
\providecommand \bibfnamefont [1]{#1}%
\providecommand \citenamefont [1]{#1}%
\providecommand \href@noop [0]{\@secondoftwo}%
\providecommand \href [0]{\begingroup \@sanitize@url \@href}%
\providecommand \@href[1]{\@@startlink{#1}\@@href}%
\providecommand \@@href[1]{\endgroup#1\@@endlink}%
\providecommand \@sanitize@url [0]{\catcode `\\12\catcode `\$12\catcode
  `\&12\catcode `\#12\catcode `\^12\catcode `\_12\catcode `\%12\relax}%
\providecommand \@@startlink[1]{}%
\providecommand \@@endlink[0]{}%
\providecommand \url  [0]{\begingroup\@sanitize@url \@url }%
\providecommand \@url [1]{\endgroup\@href {#1}{\urlprefix }}%
\providecommand \urlprefix  [0]{URL }%
\providecommand \Eprint [0]{\href }%
\providecommand \doibase [0]{http://dx.doi.org/}%
\providecommand \selectlanguage [0]{\@gobble}%
\providecommand \bibinfo  [0]{\@secondoftwo}%
\providecommand \bibfield  [0]{\@secondoftwo}%
\providecommand \translation [1]{[#1]}%
\providecommand \BibitemOpen [0]{}%
\providecommand \bibitemStop [0]{}%
\providecommand \bibitemNoStop [0]{.\EOS\space}%
\providecommand \EOS [0]{\spacefactor3000\relax}%
\providecommand \BibitemShut  [1]{\csname bibitem#1\endcsname}%
\let\auto@bib@innerbib\@empty
\bibitem [{\citenamefont {Anderson}(1995)}]{Anderson95}%
  \BibitemOpen
  \bibfield  {author} {\bibinfo {author} {\bibfnamefont {D.}~\bibnamefont
  {Anderson}},\ }\href@noop {} {\bibfield  {journal} {\bibinfo  {journal}
  {Science}\ }\textbf {\bibinfo {volume} {267}},\ \bibinfo {pages} {1618}
  (\bibinfo {year} {1995})}\BibitemShut {NoStop}%
\bibitem [{\citenamefont {Ediger}\ \emph {et~al.}(1996)\citenamefont {Ediger},
  \citenamefont {Angell},\ and\ \citenamefont {Nagel}}]{ediger1996supercooled}%
  \BibitemOpen
  \bibfield  {author} {\bibinfo {author} {\bibfnamefont {M.~D.}\ \bibnamefont
  {Ediger}}, \bibinfo {author} {\bibfnamefont {C.~A.}\ \bibnamefont {Angell}},
  \ and\ \bibinfo {author} {\bibfnamefont {S.~R.}\ \bibnamefont {Nagel}},\
  }\href@noop {} {\bibfield  {journal} {\bibinfo  {journal} {The journal of
  physical chemistry}\ }\textbf {\bibinfo {volume} {100}},\ \bibinfo {pages}
  {13200} (\bibinfo {year} {1996})}\BibitemShut {NoStop}%
\bibitem [{\citenamefont {Debenedetti}\ and\ \citenamefont
  {Stillinger}(2001)}]{Debenedetti01}%
  \BibitemOpen
  \bibfield  {author} {\bibinfo {author} {\bibfnamefont {P.}~\bibnamefont
  {Debenedetti}}\ and\ \bibinfo {author} {\bibfnamefont {F.}~\bibnamefont
  {Stillinger}},\ }\href@noop {} {\bibfield  {journal} {\bibinfo  {journal}
  {Nature}\ }\textbf {\bibinfo {volume} {410}},\ \bibinfo {pages} {259}
  (\bibinfo {year} {2001})}\BibitemShut {NoStop}%
\bibitem [{\citenamefont {Berthier}\ and\ \citenamefont
  {Biroli}(2011)}]{berthier2011theoretical}%
  \BibitemOpen
  \bibfield  {author} {\bibinfo {author} {\bibfnamefont {L.}~\bibnamefont
  {Berthier}}\ and\ \bibinfo {author} {\bibfnamefont {G.}~\bibnamefont
  {Biroli}},\ }\href@noop {} {\bibfield  {journal} {\bibinfo  {journal}
  {Reviews of modern physics}\ }\textbf {\bibinfo {volume} {83}},\ \bibinfo
  {pages} {587} (\bibinfo {year} {2011})}\BibitemShut {NoStop}%
\bibitem [{\citenamefont {Kob}\ \emph {et~al.}(1997)\citenamefont {Kob},
  \citenamefont {Donati}, \citenamefont {Plimpton}, \citenamefont {Poole},\
  and\ \citenamefont {Glotzer}}]{kob1997dynamical}%
  \BibitemOpen
  \bibfield  {author} {\bibinfo {author} {\bibfnamefont {W.}~\bibnamefont
  {Kob}}, \bibinfo {author} {\bibfnamefont {C.}~\bibnamefont {Donati}},
  \bibinfo {author} {\bibfnamefont {S.~J.}\ \bibnamefont {Plimpton}}, \bibinfo
  {author} {\bibfnamefont {P.~H.}\ \bibnamefont {Poole}}, \ and\ \bibinfo
  {author} {\bibfnamefont {S.~C.}\ \bibnamefont {Glotzer}},\ }\href@noop {}
  {\bibfield  {journal} {\bibinfo  {journal} {Physical review letters}\
  }\textbf {\bibinfo {volume} {79}},\ \bibinfo {pages} {2827} (\bibinfo {year}
  {1997})}\BibitemShut {NoStop}%
\bibitem [{\citenamefont {Yamamoto}\ and\ \citenamefont
  {Onuki}(1998)}]{yamamoto1998dynamics}%
  \BibitemOpen
  \bibfield  {author} {\bibinfo {author} {\bibfnamefont {R.}~\bibnamefont
  {Yamamoto}}\ and\ \bibinfo {author} {\bibfnamefont {A.}~\bibnamefont
  {Onuki}},\ }\href@noop {} {\bibfield  {journal} {\bibinfo  {journal}
  {Physical Review E}\ }\textbf {\bibinfo {volume} {58}},\ \bibinfo {pages}
  {3515} (\bibinfo {year} {1998})}\BibitemShut {NoStop}%
\bibitem [{\citenamefont {Dalle-Ferrier}\ \emph {et~al.}(2007)\citenamefont
  {Dalle-Ferrier}, \citenamefont {Thibierge}, \citenamefont {Alba-Simionesco},
  \citenamefont {Berthier}, \citenamefont {Biroli}, \citenamefont {Bouchaud},
  \citenamefont {Ladieu}, \citenamefont {L’H{\^o}te},\ and\ \citenamefont
  {Tarjus}}]{dalle2007spatial}%
  \BibitemOpen
  \bibfield  {author} {\bibinfo {author} {\bibfnamefont {C.}~\bibnamefont
  {Dalle-Ferrier}}, \bibinfo {author} {\bibfnamefont {C.}~\bibnamefont
  {Thibierge}}, \bibinfo {author} {\bibfnamefont {C.}~\bibnamefont
  {Alba-Simionesco}}, \bibinfo {author} {\bibfnamefont {L.}~\bibnamefont
  {Berthier}}, \bibinfo {author} {\bibfnamefont {G.}~\bibnamefont {Biroli}},
  \bibinfo {author} {\bibfnamefont {J.-P.}\ \bibnamefont {Bouchaud}}, \bibinfo
  {author} {\bibfnamefont {F.}~\bibnamefont {Ladieu}}, \bibinfo {author}
  {\bibfnamefont {D.}~\bibnamefont {L’H{\^o}te}}, \ and\ \bibinfo {author}
  {\bibfnamefont {G.}~\bibnamefont {Tarjus}},\ }\href@noop {} {\bibfield
  {journal} {\bibinfo  {journal} {Phys. Rev. E}\ }\textbf {\bibinfo {volume}
  {76}},\ \bibinfo {pages} {041510} (\bibinfo {year} {2007})}\BibitemShut
  {NoStop}%
\bibitem [{\citenamefont {Karmakar}\ \emph {et~al.}(2014)\citenamefont
  {Karmakar}, \citenamefont {Dasgupta},\ and\ \citenamefont
  {Sastry}}]{karmakar2014growing}%
  \BibitemOpen
  \bibfield  {author} {\bibinfo {author} {\bibfnamefont {S.}~\bibnamefont
  {Karmakar}}, \bibinfo {author} {\bibfnamefont {C.}~\bibnamefont {Dasgupta}},
  \ and\ \bibinfo {author} {\bibfnamefont {S.}~\bibnamefont {Sastry}},\
  }\href@noop {} {\bibfield  {journal} {\bibinfo  {journal} {Annu. Rev.
  Condens. Matter Phys.}\ }\textbf {\bibinfo {volume} {5}},\ \bibinfo {pages}
  {255} (\bibinfo {year} {2014})}\BibitemShut {NoStop}%
\bibitem [{\citenamefont {Adam}\ and\ \citenamefont
  {Gibbs}(1965)}]{adam1965temperature}%
  \BibitemOpen
  \bibfield  {author} {\bibinfo {author} {\bibfnamefont {G.}~\bibnamefont
  {Adam}}\ and\ \bibinfo {author} {\bibfnamefont {J.~H.}\ \bibnamefont
  {Gibbs}},\ }\href@noop {} {\bibfield  {journal} {\bibinfo  {journal} {The
  journal of chemical physics}\ }\textbf {\bibinfo {volume} {43}},\ \bibinfo
  {pages} {139} (\bibinfo {year} {1965})}\BibitemShut {NoStop}%
\bibitem [{\citenamefont {Kirkpatrick}\ \emph {et~al.}(1989)\citenamefont
  {Kirkpatrick}, \citenamefont {Thirumalai},\ and\ \citenamefont
  {Wolynes}}]{kirkpatrick1989scaling}%
  \BibitemOpen
  \bibfield  {author} {\bibinfo {author} {\bibfnamefont {T.~R.}\ \bibnamefont
  {Kirkpatrick}}, \bibinfo {author} {\bibfnamefont {D.}~\bibnamefont
  {Thirumalai}}, \ and\ \bibinfo {author} {\bibfnamefont {P.~G.}\ \bibnamefont
  {Wolynes}},\ }\href@noop {} {\bibfield  {journal} {\bibinfo  {journal}
  {Physical Review A}\ }\textbf {\bibinfo {volume} {40}},\ \bibinfo {pages}
  {1045} (\bibinfo {year} {1989})}\BibitemShut {NoStop}%
\bibitem [{\citenamefont {Lubchenko}\ and\ \citenamefont
  {Wolynes}(2007)}]{lubchenko2007theory}%
  \BibitemOpen
  \bibfield  {author} {\bibinfo {author} {\bibfnamefont {V.}~\bibnamefont
  {Lubchenko}}\ and\ \bibinfo {author} {\bibfnamefont {P.~G.}\ \bibnamefont
  {Wolynes}},\ }\href@noop {} {\bibfield  {journal} {\bibinfo  {journal} {Annu.
  Rev. Phys. Chem.}\ }\textbf {\bibinfo {volume} {58}},\ \bibinfo {pages} {235}
  (\bibinfo {year} {2007})}\BibitemShut {NoStop}%
\bibitem [{\citenamefont {Biroli}\ and\ \citenamefont
  {Bouchaud}(2012)}]{Biroli12}%
  \BibitemOpen
  \bibfield  {author} {\bibinfo {author} {\bibfnamefont {G.}~\bibnamefont
  {Biroli}}\ and\ \bibinfo {author} {\bibfnamefont {J.-P.}\ \bibnamefont
  {Bouchaud}},\ }\href@noop {} {\bibfield  {journal} {\bibinfo  {journal}
  {Structural Glasses and Supercooled Liquids: Theory, Experiment, and
  Applications}\ ,\ \bibinfo {pages} {31}} (\bibinfo {year}
  {2012})}\BibitemShut {NoStop}%
\bibitem [{\citenamefont {Patrick~Royall}\ \emph {et~al.}(2008)\citenamefont
  {Patrick~Royall}, \citenamefont {Williams}, \citenamefont {Ohtsuka},\ and\
  \citenamefont {Tanaka}}]{patrick2008direct}%
  \BibitemOpen
  \bibfield  {author} {\bibinfo {author} {\bibfnamefont {C.}~\bibnamefont
  {Patrick~Royall}}, \bibinfo {author} {\bibfnamefont {S.~R.}\ \bibnamefont
  {Williams}}, \bibinfo {author} {\bibfnamefont {T.}~\bibnamefont {Ohtsuka}}, \
  and\ \bibinfo {author} {\bibfnamefont {H.}~\bibnamefont {Tanaka}},\
  }\href@noop {} {\bibfield  {journal} {\bibinfo  {journal} {Nature materials}\
  }\textbf {\bibinfo {volume} {7}},\ \bibinfo {pages} {556} (\bibinfo {year}
  {2008})}\BibitemShut {NoStop}%
\bibitem [{\citenamefont {Kawasaki}\ \emph {et~al.}(2007)\citenamefont
  {Kawasaki}, \citenamefont {Araki},\ and\ \citenamefont
  {Tanaka}}]{kawasaki2007correlation}%
  \BibitemOpen
  \bibfield  {author} {\bibinfo {author} {\bibfnamefont {T.}~\bibnamefont
  {Kawasaki}}, \bibinfo {author} {\bibfnamefont {T.}~\bibnamefont {Araki}}, \
  and\ \bibinfo {author} {\bibfnamefont {H.}~\bibnamefont {Tanaka}},\
  }\href@noop {} {\bibfield  {journal} {\bibinfo  {journal} {Physical review
  letters}\ }\textbf {\bibinfo {volume} {99}},\ \bibinfo {pages} {215701}
  (\bibinfo {year} {2007})}\BibitemShut {NoStop}%
\bibitem [{\citenamefont {Ritort}\ and\ \citenamefont
  {Sollich}(2003)}]{ritort2003glassy}%
  \BibitemOpen
  \bibfield  {author} {\bibinfo {author} {\bibfnamefont {F.}~\bibnamefont
  {Ritort}}\ and\ \bibinfo {author} {\bibfnamefont {P.}~\bibnamefont
  {Sollich}},\ }\href@noop {} {\bibfield  {journal} {\bibinfo  {journal}
  {Advances in physics}\ }\textbf {\bibinfo {volume} {52}},\ \bibinfo {pages}
  {219} (\bibinfo {year} {2003})}\BibitemShut {NoStop}%
\bibitem [{\citenamefont {Berthier}\ \emph {et~al.}(2011)\citenamefont
  {Berthier}, \citenamefont {Biroli}, \citenamefont {Bouchaud}, \citenamefont
  {Cipelletti},\ and\ \citenamefont {van Saarloos}}]{berthier2011dynamical}%
  \BibitemOpen
  \bibfield  {author} {\bibinfo {author} {\bibfnamefont {L.}~\bibnamefont
  {Berthier}}, \bibinfo {author} {\bibfnamefont {G.}~\bibnamefont {Biroli}},
  \bibinfo {author} {\bibfnamefont {J.-P.}\ \bibnamefont {Bouchaud}}, \bibinfo
  {author} {\bibfnamefont {L.}~\bibnamefont {Cipelletti}}, \ and\ \bibinfo
  {author} {\bibfnamefont {W.}~\bibnamefont {van Saarloos}},\ }\href@noop {}
  {\emph {\bibinfo {title} {Dynamical heterogeneities in glasses, colloids, and
  granular media}}},\ Vol.\ \bibinfo {volume} {150}\ (\bibinfo  {publisher}
  {OUP Oxford},\ \bibinfo {year} {2011})\BibitemShut {NoStop}%
\bibitem [{\citenamefont {Garrahan}\ and\ \citenamefont
  {Chandler}(2002)}]{garrahan2002geometrical}%
  \BibitemOpen
  \bibfield  {author} {\bibinfo {author} {\bibfnamefont {J.~P.}\ \bibnamefont
  {Garrahan}}\ and\ \bibinfo {author} {\bibfnamefont {D.}~\bibnamefont
  {Chandler}},\ }\href@noop {} {\bibfield  {journal} {\bibinfo  {journal}
  {Physical review letters}\ }\textbf {\bibinfo {volume} {89}},\ \bibinfo
  {pages} {035704} (\bibinfo {year} {2002})}\BibitemShut {NoStop}%
\bibitem [{\citenamefont {Garrahan}\ \emph {et~al.}(2007)\citenamefont
  {Garrahan}, \citenamefont {Jack}, \citenamefont {Lecomte}, \citenamefont
  {Pitard}, \citenamefont {van Duijvendijk},\ and\ \citenamefont {van
  Wijland}}]{garrahan2007dynamical}%
  \BibitemOpen
  \bibfield  {author} {\bibinfo {author} {\bibfnamefont {J.~P.}\ \bibnamefont
  {Garrahan}}, \bibinfo {author} {\bibfnamefont {R.~L.}\ \bibnamefont {Jack}},
  \bibinfo {author} {\bibfnamefont {V.}~\bibnamefont {Lecomte}}, \bibinfo
  {author} {\bibfnamefont {E.}~\bibnamefont {Pitard}}, \bibinfo {author}
  {\bibfnamefont {K.}~\bibnamefont {van Duijvendijk}}, \ and\ \bibinfo {author}
  {\bibfnamefont {F.}~\bibnamefont {van Wijland}},\ }\href@noop {} {\bibfield
  {journal} {\bibinfo  {journal} {Physical review letters}\ }\textbf {\bibinfo
  {volume} {98}},\ \bibinfo {pages} {195702} (\bibinfo {year}
  {2007})}\BibitemShut {NoStop}%
\bibitem [{\citenamefont {Hedges}\ \emph {et~al.}(2009)\citenamefont {Hedges},
  \citenamefont {Jack}, \citenamefont {Garrahan},\ and\ \citenamefont
  {Chandler}}]{hedges2009dynamic}%
  \BibitemOpen
  \bibfield  {author} {\bibinfo {author} {\bibfnamefont {L.}~\bibnamefont
  {Hedges}}, \bibinfo {author} {\bibfnamefont {R.}~\bibnamefont {Jack}},
  \bibinfo {author} {\bibfnamefont {J.}~\bibnamefont {Garrahan}}, \ and\
  \bibinfo {author} {\bibfnamefont {D.}~\bibnamefont {Chandler}},\ }\href@noop
  {} {\bibfield  {journal} {\bibinfo  {journal} {Science}\ }\textbf {\bibinfo
  {volume} {323}},\ \bibinfo {pages} {1309} (\bibinfo {year}
  {2009})}\BibitemShut {NoStop}%
\bibitem [{\citenamefont {Turnbull}\ and\ \citenamefont
  {Cohen}(1961)}]{turnbull1961free}%
  \BibitemOpen
  \bibfield  {author} {\bibinfo {author} {\bibfnamefont {D.}~\bibnamefont
  {Turnbull}}\ and\ \bibinfo {author} {\bibfnamefont {M.~H.}\ \bibnamefont
  {Cohen}},\ }\href@noop {} {\bibfield  {journal} {\bibinfo  {journal} {The
  Journal of chemical physics}\ }\textbf {\bibinfo {volume} {34}},\ \bibinfo
  {pages} {120} (\bibinfo {year} {1961})}\BibitemShut {NoStop}%
\bibitem [{\citenamefont {Dyre}(2006)}]{dyre2006colloquium}%
  \BibitemOpen
  \bibfield  {author} {\bibinfo {author} {\bibfnamefont {J.}~\bibnamefont
  {Dyre}},\ }\href@noop {} {\bibfield  {journal} {\bibinfo  {journal} {Rev.
  Mod. Phys.}\ }\textbf {\bibinfo {volume} {78}},\ \bibinfo {pages} {953}
  (\bibinfo {year} {2006})}\BibitemShut {NoStop}%
\bibitem [{\citenamefont {Kapteijns}\ \emph {et~al.}(2021)\citenamefont
  {Kapteijns}, \citenamefont {Richard}, \citenamefont {Bouchbinder},
  \citenamefont {Schr{\o}der}, \citenamefont {Dyre},\ and\ \citenamefont
  {Lerner}}]{JeppeEdan}%
  \BibitemOpen
  \bibfield  {author} {\bibinfo {author} {\bibfnamefont {G.}~\bibnamefont
  {Kapteijns}}, \bibinfo {author} {\bibfnamefont {D.}~\bibnamefont {Richard}},
  \bibinfo {author} {\bibfnamefont {E.}~\bibnamefont {Bouchbinder}}, \bibinfo
  {author} {\bibfnamefont {T.~B.}\ \bibnamefont {Schr{\o}der}}, \bibinfo
  {author} {\bibfnamefont {J.~C.}\ \bibnamefont {Dyre}}, \ and\ \bibinfo
  {author} {\bibfnamefont {E.}~\bibnamefont {Lerner}},\ }\href {\doibase
  10.1063/5.0051193} {\bibfield  {journal} {\bibinfo  {journal} {The Journal of
  Chemical Physics}\ }\textbf {\bibinfo {volume} {155}},\ \bibinfo {pages}
  {74502} (\bibinfo {year} {2021})}\BibitemShut {NoStop}%
\bibitem [{\citenamefont {Lerbinger}\ \emph {et~al.}(2022)\citenamefont
  {Lerbinger}, \citenamefont {Barbot}, \citenamefont {Vandembroucq},\ and\
  \citenamefont {Patinet}}]{lerbinger2022relevance}%
  \BibitemOpen
  \bibfield  {author} {\bibinfo {author} {\bibfnamefont {M.}~\bibnamefont
  {Lerbinger}}, \bibinfo {author} {\bibfnamefont {A.}~\bibnamefont {Barbot}},
  \bibinfo {author} {\bibfnamefont {D.}~\bibnamefont {Vandembroucq}}, \ and\
  \bibinfo {author} {\bibfnamefont {S.}~\bibnamefont {Patinet}},\ }\href@noop
  {} {\bibfield  {journal} {\bibinfo  {journal} {Physical Review Letters}\
  }\textbf {\bibinfo {volume} {129}},\ \bibinfo {pages} {195501} (\bibinfo
  {year} {2022})}\BibitemShut {NoStop}%
\bibitem [{\citenamefont {Ciamarra}\ \emph {et~al.}(2023)\citenamefont
  {Ciamarra}, \citenamefont {Ji},\ and\ \citenamefont {Wyart}}]{massimo}%
  \BibitemOpen
  \bibfield  {author} {\bibinfo {author} {\bibfnamefont {M.~P.}\ \bibnamefont
  {Ciamarra}}, \bibinfo {author} {\bibfnamefont {W.}~\bibnamefont {Ji}}, \ and\
  \bibinfo {author} {\bibfnamefont {M.}~\bibnamefont {Wyart}},\ }\href
  {\doibase 10.48550/ARXIV.2302.05150} {\enquote {\bibinfo {title} {The energy
  cost of local rearrangements, not cooperative effects, makes glasses
  solid},}\ } (\bibinfo {year} {2023})\BibitemShut {NoStop}%
\bibitem [{\citenamefont {Lema{\^{i}}tre}(2014)}]{Lemaitre14}%
  \BibitemOpen
  \bibfield  {author} {\bibinfo {author} {\bibfnamefont {A.}~\bibnamefont
  {Lema{\^{i}}tre}},\ }\href@noop {} {\bibfield  {journal} {\bibinfo  {journal}
  {Phys. Rev. Lett.}\ }\textbf {\bibinfo {volume} {113}},\ \bibinfo {pages}
  {245702} (\bibinfo {year} {2014})}\BibitemShut {NoStop}%
\bibitem [{\citenamefont {Chowdhury}\ \emph {et~al.}(2016)\citenamefont
  {Chowdhury}, \citenamefont {Abraham}, \citenamefont {Hudson},\ and\
  \citenamefont {Harrowell}}]{chowdhury2016long}%
  \BibitemOpen
  \bibfield  {author} {\bibinfo {author} {\bibfnamefont {S.}~\bibnamefont
  {Chowdhury}}, \bibinfo {author} {\bibfnamefont {S.}~\bibnamefont {Abraham}},
  \bibinfo {author} {\bibfnamefont {T.}~\bibnamefont {Hudson}}, \ and\ \bibinfo
  {author} {\bibfnamefont {P.}~\bibnamefont {Harrowell}},\ }\href@noop {}
  {\bibfield  {journal} {\bibinfo  {journal} {The Journal of chemical physics}\
  }\textbf {\bibinfo {volume} {144}},\ \bibinfo {pages} {124508} (\bibinfo
  {year} {2016})}\BibitemShut {NoStop}%
\bibitem [{\citenamefont {Tong}\ \emph {et~al.}(2020)\citenamefont {Tong},
  \citenamefont {Sengupta},\ and\ \citenamefont {Tanaka}}]{tong2020emergent}%
  \BibitemOpen
  \bibfield  {author} {\bibinfo {author} {\bibfnamefont {H.}~\bibnamefont
  {Tong}}, \bibinfo {author} {\bibfnamefont {S.}~\bibnamefont {Sengupta}}, \
  and\ \bibinfo {author} {\bibfnamefont {H.}~\bibnamefont {Tanaka}},\
  }\href@noop {} {\bibfield  {journal} {\bibinfo  {journal} {Nature
  communications}\ }\textbf {\bibinfo {volume} {11}},\ \bibinfo {pages} {1}
  (\bibinfo {year} {2020})}\BibitemShut {NoStop}%
\bibitem [{\citenamefont {Wu}\ \emph {et~al.}(2015)\citenamefont {Wu},
  \citenamefont {Iwashita},\ and\ \citenamefont {Egami}}]{wu2015anisotropic}%
  \BibitemOpen
  \bibfield  {author} {\bibinfo {author} {\bibfnamefont {B.}~\bibnamefont
  {Wu}}, \bibinfo {author} {\bibfnamefont {T.}~\bibnamefont {Iwashita}}, \ and\
  \bibinfo {author} {\bibfnamefont {T.}~\bibnamefont {Egami}},\ }\href@noop {}
  {\bibfield  {journal} {\bibinfo  {journal} {Physical Review E}\ }\textbf
  {\bibinfo {volume} {91}},\ \bibinfo {pages} {032301} (\bibinfo {year}
  {2015})}\BibitemShut {NoStop}%
\bibitem [{\citenamefont {Maier}\ \emph {et~al.}(2017)\citenamefont {Maier},
  \citenamefont {Zippelius},\ and\ \citenamefont {Fuchs}}]{maier2017emergence}%
  \BibitemOpen
  \bibfield  {author} {\bibinfo {author} {\bibfnamefont {M.}~\bibnamefont
  {Maier}}, \bibinfo {author} {\bibfnamefont {A.}~\bibnamefont {Zippelius}}, \
  and\ \bibinfo {author} {\bibfnamefont {M.}~\bibnamefont {Fuchs}},\
  }\href@noop {} {\bibfield  {journal} {\bibinfo  {journal} {Physical review
  letters}\ }\textbf {\bibinfo {volume} {119}},\ \bibinfo {pages} {265701}
  (\bibinfo {year} {2017})}\BibitemShut {NoStop}%
\bibitem [{\citenamefont {Steffen}\ \emph {et~al.}(2022)\citenamefont
  {Steffen}, \citenamefont {Schneider}, \citenamefont {M{\"u}ller},\ and\
  \citenamefont {Rottler}}]{steffen2022molecular}%
  \BibitemOpen
  \bibfield  {author} {\bibinfo {author} {\bibfnamefont {D.}~\bibnamefont
  {Steffen}}, \bibinfo {author} {\bibfnamefont {L.}~\bibnamefont {Schneider}},
  \bibinfo {author} {\bibfnamefont {M.}~\bibnamefont {M{\"u}ller}}, \ and\
  \bibinfo {author} {\bibfnamefont {J.}~\bibnamefont {Rottler}},\ }\href@noop
  {} {\bibfield  {journal} {\bibinfo  {journal} {The Journal of Chemical
  Physics}\ }\textbf {\bibinfo {volume} {157}},\ \bibinfo {pages} {064501}
  (\bibinfo {year} {2022})}\BibitemShut {NoStop}%
\bibitem [{\citenamefont {Flenner}\ and\ \citenamefont
  {Szamel}(2015)}]{flenner2015long}%
  \BibitemOpen
  \bibfield  {author} {\bibinfo {author} {\bibfnamefont {E.}~\bibnamefont
  {Flenner}}\ and\ \bibinfo {author} {\bibfnamefont {G.}~\bibnamefont
  {Szamel}},\ }\href@noop {} {\bibfield  {journal} {\bibinfo  {journal}
  {Physical Review Letters}\ }\textbf {\bibinfo {volume} {114}},\ \bibinfo
  {pages} {025501} (\bibinfo {year} {2015})}\BibitemShut {NoStop}%
\bibitem [{\citenamefont {Klochko}\ \emph {et~al.}(2022)\citenamefont
  {Klochko}, \citenamefont {Baschnagel}, \citenamefont {Wittmer}, \citenamefont
  {Meyer}, \citenamefont {Benzerara},\ and\ \citenamefont
  {Semenov}}]{klochko2022theory}%
  \BibitemOpen
  \bibfield  {author} {\bibinfo {author} {\bibfnamefont {L.}~\bibnamefont
  {Klochko}}, \bibinfo {author} {\bibfnamefont {J.}~\bibnamefont {Baschnagel}},
  \bibinfo {author} {\bibfnamefont {J.}~\bibnamefont {Wittmer}}, \bibinfo
  {author} {\bibfnamefont {H.}~\bibnamefont {Meyer}}, \bibinfo {author}
  {\bibfnamefont {O.}~\bibnamefont {Benzerara}}, \ and\ \bibinfo {author}
  {\bibfnamefont {A.}~\bibnamefont {Semenov}},\ }\href@noop {} {\bibfield
  {journal} {\bibinfo  {journal} {The Journal of Chemical Physics}\ }\textbf
  {\bibinfo {volume} {156}},\ \bibinfo {pages} {164505} (\bibinfo {year}
  {2022})}\BibitemShut {NoStop}%
\bibitem [{\citenamefont {Chacko}\ \emph {et~al.}(2021)\citenamefont {Chacko},
  \citenamefont {Landes}, \citenamefont {Biroli}, \citenamefont {Dauchot},
  \citenamefont {Liu},\ and\ \citenamefont
  {Reichman}}]{chacko2021elastoplasticity}%
  \BibitemOpen
  \bibfield  {author} {\bibinfo {author} {\bibfnamefont {R.}~\bibnamefont
  {Chacko}}, \bibinfo {author} {\bibfnamefont {F.}~\bibnamefont {Landes}},
  \bibinfo {author} {\bibfnamefont {G.}~\bibnamefont {Biroli}}, \bibinfo
  {author} {\bibfnamefont {O.}~\bibnamefont {Dauchot}}, \bibinfo {author}
  {\bibfnamefont {A.}~\bibnamefont {Liu}}, \ and\ \bibinfo {author}
  {\bibfnamefont {D.}~\bibnamefont {Reichman}},\ }\href {\doibase
  10.48550/arXiv.2103.01852} {\bibfield  {journal} {\bibinfo  {journal} {arXiv
  preprint 2103.01852}\ } (\bibinfo {year} {2021}),\
  10.48550/arXiv.2103.01852}\BibitemShut {NoStop}%
\bibitem [{\citenamefont {Ozawa}\ and\ \citenamefont
  {Biroli}(2023)}]{ozawa2023elasticity}%
  \BibitemOpen
  \bibfield  {author} {\bibinfo {author} {\bibfnamefont {M.}~\bibnamefont
  {Ozawa}}\ and\ \bibinfo {author} {\bibfnamefont {G.}~\bibnamefont {Biroli}},\
  }\href@noop {} {\bibfield  {journal} {\bibinfo  {journal} {Physical Review
  Letters}\ }\textbf {\bibinfo {volume} {130}},\ \bibinfo {pages} {138201}
  (\bibinfo {year} {2023})}\BibitemShut {NoStop}%
\bibitem [{\citenamefont {Tahaei}\ \emph {et~al.}(2023)\citenamefont {Tahaei},
  \citenamefont {Biroli}, \citenamefont {Ozawa}, \citenamefont {Popovi{\'c}},\
  and\ \citenamefont {Wyart}}]{tahaei2023scaling}%
  \BibitemOpen
  \bibfield  {author} {\bibinfo {author} {\bibfnamefont {A.}~\bibnamefont
  {Tahaei}}, \bibinfo {author} {\bibfnamefont {G.}~\bibnamefont {Biroli}},
  \bibinfo {author} {\bibfnamefont {M.}~\bibnamefont {Ozawa}}, \bibinfo
  {author} {\bibfnamefont {M.}~\bibnamefont {Popovi{\'c}}}, \ and\ \bibinfo
  {author} {\bibfnamefont {M.}~\bibnamefont {Wyart}},\ }\href@noop {}
  {\bibfield  {journal} {\bibinfo  {journal} {arXiv preprint arXiv:2305.00219}\
  } (\bibinfo {year} {2023})}\BibitemShut {NoStop}%
\bibitem [{\citenamefont {Kob}(1999)}]{kob1999computer}%
  \BibitemOpen
  \bibfield  {author} {\bibinfo {author} {\bibfnamefont {W.}~\bibnamefont
  {Kob}},\ }\href@noop {} {\bibfield  {journal} {\bibinfo  {journal} {Journal
  of Physics: Condensed Matter}\ }\textbf {\bibinfo {volume} {11}},\ \bibinfo
  {pages} {R85} (\bibinfo {year} {1999})}\BibitemShut {NoStop}%
\bibitem [{\citenamefont {Briano}\ and\ \citenamefont
  {Glandt}(1984)}]{Glandt84}%
  \BibitemOpen
  \bibfield  {author} {\bibinfo {author} {\bibfnamefont {J.}~\bibnamefont
  {Briano}}\ and\ \bibinfo {author} {\bibfnamefont {E.}~\bibnamefont
  {Glandt}},\ }\href {\doibase 10.1063/1.447087} {\bibfield  {journal}
  {\bibinfo  {journal} {J. Chem. Phys.}\ }\textbf {\bibinfo {volume} {80}},\
  \bibinfo {pages} {3336} (\bibinfo {year} {1984})}\BibitemShut {NoStop}%
\bibitem [{\citenamefont {Grigera}\ and\ \citenamefont
  {Parisi}(2001)}]{Grigera01fast}%
  \BibitemOpen
  \bibfield  {author} {\bibinfo {author} {\bibfnamefont {T.}~\bibnamefont
  {Grigera}}\ and\ \bibinfo {author} {\bibfnamefont {G.}~\bibnamefont
  {Parisi}},\ }\href@noop {} {\bibfield  {journal} {\bibinfo  {journal} {Phys.
  Rev. E}\ }\textbf {\bibinfo {volume} {63}},\ \bibinfo {pages} {045102}
  (\bibinfo {year} {2001})}\BibitemShut {NoStop}%
\bibitem [{\citenamefont {Fern{\'a}ndez}\ \emph {et~al.}(2007)\citenamefont
  {Fern{\'a}ndez}, \citenamefont {Mart{\'\i}n-Mayor},\ and\ \citenamefont
  {Verrocchio}}]{fernandez2007phase}%
  \BibitemOpen
  \bibfield  {author} {\bibinfo {author} {\bibfnamefont {L.}~\bibnamefont
  {Fern{\'a}ndez}}, \bibinfo {author} {\bibfnamefont {V.}~\bibnamefont
  {Mart{\'\i}n-Mayor}}, \ and\ \bibinfo {author} {\bibfnamefont
  {P.}~\bibnamefont {Verrocchio}},\ }\href@noop {} {\bibfield  {journal}
  {\bibinfo  {journal} {Physical review letters}\ }\textbf {\bibinfo {volume}
  {98}},\ \bibinfo {pages} {085702} (\bibinfo {year} {2007})}\BibitemShut
  {NoStop}%
\bibitem [{\citenamefont {Guti{\'e}rrez}\ \emph {et~al.}(2015)\citenamefont
  {Guti{\'e}rrez}, \citenamefont {Karmakar}, \citenamefont {Pollack},\ and\
  \citenamefont {Procaccia}}]{gutierrez2015static}%
  \BibitemOpen
  \bibfield  {author} {\bibinfo {author} {\bibfnamefont {R.}~\bibnamefont
  {Guti{\'e}rrez}}, \bibinfo {author} {\bibfnamefont {S.}~\bibnamefont
  {Karmakar}}, \bibinfo {author} {\bibfnamefont {Y.}~\bibnamefont {Pollack}}, \
  and\ \bibinfo {author} {\bibfnamefont {I.}~\bibnamefont {Procaccia}},\
  }\href@noop {} {\bibfield  {journal} {\bibinfo  {journal} {Europhys. Lett.}\
  }\textbf {\bibinfo {volume} {111}},\ \bibinfo {pages} {56009} (\bibinfo
  {year} {2015})}\BibitemShut {NoStop}%
\bibitem [{\citenamefont {Ninarello}\ \emph {et~al.}(2017)\citenamefont
  {Ninarello}, \citenamefont {Berthier},\ and\ \citenamefont
  {Coslovich}}]{ninarello2017models}%
  \BibitemOpen
  \bibfield  {author} {\bibinfo {author} {\bibfnamefont {A.}~\bibnamefont
  {Ninarello}}, \bibinfo {author} {\bibfnamefont {L.}~\bibnamefont {Berthier}},
  \ and\ \bibinfo {author} {\bibfnamefont {D.}~\bibnamefont {Coslovich}},\
  }\href@noop {} {\bibfield  {journal} {\bibinfo  {journal} {Phys. Rev. X}\
  }\textbf {\bibinfo {volume} {7}},\ \bibinfo {pages} {021039} (\bibinfo {year}
  {2017})}\BibitemShut {NoStop}%
\bibitem [{\citenamefont {Brito}\ \emph {et~al.}(2018)\citenamefont {Brito},
  \citenamefont {Lerner},\ and\ \citenamefont {Wyart}}]{Brito18}%
  \BibitemOpen
  \bibfield  {author} {\bibinfo {author} {\bibfnamefont {C.}~\bibnamefont
  {Brito}}, \bibinfo {author} {\bibfnamefont {E.}~\bibnamefont {Lerner}}, \
  and\ \bibinfo {author} {\bibfnamefont {M.}~\bibnamefont {Wyart}},\ }\href
  {\doibase 10.1103/PhysRevX.8.031050} {\bibfield  {journal} {\bibinfo
  {journal} {Phys. Rev. X}\ }\textbf {\bibinfo {volume} {8}},\ \bibinfo {pages}
  {031050} (\bibinfo {year} {2018})}\BibitemShut {NoStop}%
\bibitem [{\citenamefont {Hagh}\ \emph {et~al.}(2022)\citenamefont {Hagh},
  \citenamefont {Nagel}, \citenamefont {Liu}, \citenamefont {Manning},\ and\
  \citenamefont {Corwin}}]{hagh2021transient}%
  \BibitemOpen
  \bibfield  {author} {\bibinfo {author} {\bibfnamefont {V.~F.}\ \bibnamefont
  {Hagh}}, \bibinfo {author} {\bibfnamefont {S.~R.}\ \bibnamefont {Nagel}},
  \bibinfo {author} {\bibfnamefont {A.~J.}\ \bibnamefont {Liu}}, \bibinfo
  {author} {\bibfnamefont {M.~L.}\ \bibnamefont {Manning}}, \ and\ \bibinfo
  {author} {\bibfnamefont {E.~I.}\ \bibnamefont {Corwin}},\ }\href@noop {}
  {\bibfield  {journal} {\bibinfo  {journal} {Proceedings of the National
  Academy of Sciences}\ }\textbf {\bibinfo {volume} {119}},\ \bibinfo {pages}
  {e2117622119} (\bibinfo {year} {2022})}\BibitemShut {NoStop}%
\bibitem [{\citenamefont {Szamel}(2019)}]{szamel2019theory}%
  \BibitemOpen
  \bibfield  {author} {\bibinfo {author} {\bibfnamefont {G.}~\bibnamefont
  {Szamel}},\ }\href@noop {} {\bibfield  {journal} {\bibinfo  {journal}
  {Journal of Statistical Mechanics: Theory and Experiment}\ }\textbf {\bibinfo
  {volume} {2019}},\ \bibinfo {pages} {104016} (\bibinfo {year}
  {2019})}\BibitemShut {NoStop}%
\bibitem [{\citenamefont {Ikeda}\ \emph {et~al.}(2019)\citenamefont {Ikeda},
  \citenamefont {Urbani},\ and\ \citenamefont {Zamponi}}]{ikeda2019mean}%
  \BibitemOpen
  \bibfield  {author} {\bibinfo {author} {\bibfnamefont {H.}~\bibnamefont
  {Ikeda}}, \bibinfo {author} {\bibfnamefont {P.}~\bibnamefont {Urbani}}, \
  and\ \bibinfo {author} {\bibfnamefont {F.}~\bibnamefont {Zamponi}},\
  }\href@noop {} {\bibfield  {journal} {\bibinfo  {journal} {Journal of Physics
  A: Mathematical and Theoretical}\ }\textbf {\bibinfo {volume} {52}},\
  \bibinfo {pages} {344001} (\bibinfo {year} {2019})}\BibitemShut {NoStop}%
\bibitem [{\citenamefont {Wyart}\ and\ \citenamefont {Cates}(2017)}]{Wyart17}%
  \BibitemOpen
  \bibfield  {author} {\bibinfo {author} {\bibfnamefont {M.}~\bibnamefont
  {Wyart}}\ and\ \bibinfo {author} {\bibfnamefont {M.}~\bibnamefont {Cates}},\
  }\href {\doibase 10.1103/PhysRevLett.119.195501} {\bibfield  {journal}
  {\bibinfo  {journal} {Phys. Rev. Lett.}\ }\textbf {\bibinfo {volume} {119}},\
  \bibinfo {pages} {195501} (\bibinfo {year} {2017})}\BibitemShut {NoStop}%
\bibitem [{\citenamefont {Gopinath}\ \emph {et~al.}(2022)\citenamefont
  {Gopinath}, \citenamefont {Lee}, \citenamefont {Gao}, \citenamefont {An},
  \citenamefont {Chan}, \citenamefont {Yip}, \citenamefont {Deng},\ and\
  \citenamefont {Lam}}]{gopinath2022diffusion}%
  \BibitemOpen
  \bibfield  {author} {\bibinfo {author} {\bibfnamefont {G.}~\bibnamefont
  {Gopinath}}, \bibinfo {author} {\bibfnamefont {C.-S.}\ \bibnamefont {Lee}},
  \bibinfo {author} {\bibfnamefont {X.-Y.}\ \bibnamefont {Gao}}, \bibinfo
  {author} {\bibfnamefont {X.-D.}\ \bibnamefont {An}}, \bibinfo {author}
  {\bibfnamefont {C.-H.}\ \bibnamefont {Chan}}, \bibinfo {author}
  {\bibfnamefont {C.-T.}\ \bibnamefont {Yip}}, \bibinfo {author} {\bibfnamefont
  {H.-Y.}\ \bibnamefont {Deng}}, \ and\ \bibinfo {author} {\bibfnamefont
  {C.-H.}\ \bibnamefont {Lam}},\ }\href@noop {} {\bibfield  {journal} {\bibinfo
   {journal} {Physical Review Letters}\ }\textbf {\bibinfo {volume} {129}},\
  \bibinfo {pages} {168002} (\bibinfo {year} {2022})}\BibitemShut {NoStop}%
\bibitem [{\citenamefont {Cammarota}\ and\ \citenamefont
  {Biroli}(2012)}]{cammarota2012ideal}%
  \BibitemOpen
  \bibfield  {author} {\bibinfo {author} {\bibfnamefont {C.}~\bibnamefont
  {Cammarota}}\ and\ \bibinfo {author} {\bibfnamefont {G.}~\bibnamefont
  {Biroli}},\ }\href@noop {} {\bibfield  {journal} {\bibinfo  {journal}
  {Proceedings of the National Academy of Sciences}\ }\textbf {\bibinfo
  {volume} {109}},\ \bibinfo {pages} {8850} (\bibinfo {year}
  {2012})}\BibitemShut {NoStop}%
\bibitem [{\citenamefont {Berthier}\ \emph {et~al.}(2016)\citenamefont
  {Berthier}, \citenamefont {Coslovich}, \citenamefont {Ninarello},\ and\
  \citenamefont {Ozawa}}]{berthier2016equilibrium}%
  \BibitemOpen
  \bibfield  {author} {\bibinfo {author} {\bibfnamefont {L.}~\bibnamefont
  {Berthier}}, \bibinfo {author} {\bibfnamefont {D.}~\bibnamefont {Coslovich}},
  \bibinfo {author} {\bibfnamefont {A.}~\bibnamefont {Ninarello}}, \ and\
  \bibinfo {author} {\bibfnamefont {M.}~\bibnamefont {Ozawa}},\ }\href@noop {}
  {\bibfield  {journal} {\bibinfo  {journal} {Physical review letters}\
  }\textbf {\bibinfo {volume} {116}},\ \bibinfo {pages} {238002} (\bibinfo
  {year} {2016})}\BibitemShut {NoStop}%
\bibitem [{\citenamefont {Biroli}\ and\ \citenamefont
  {Bouchaud}(2023)}]{biroli2023rfot}%
  \BibitemOpen
  \bibfield  {author} {\bibinfo {author} {\bibfnamefont {G.}~\bibnamefont
  {Biroli}}\ and\ \bibinfo {author} {\bibfnamefont {J.-P.}\ \bibnamefont
  {Bouchaud}},\ }\href@noop {} {\bibfield  {journal} {\bibinfo  {journal}
  {Comptes Rendus. Physique}\ }\textbf {\bibinfo {volume} {24}},\ \bibinfo
  {pages} {1} (\bibinfo {year} {2023})}\BibitemShut {NoStop}%
\bibitem [{\citenamefont {Chandler}\ and\ \citenamefont
  {Garrahan}(2010)}]{chandler2010dynamics}%
  \BibitemOpen
  \bibfield  {author} {\bibinfo {author} {\bibfnamefont {D.}~\bibnamefont
  {Chandler}}\ and\ \bibinfo {author} {\bibfnamefont {J.~P.}\ \bibnamefont
  {Garrahan}},\ }\href@noop {} {\bibfield  {journal} {\bibinfo  {journal}
  {Annual review of physical chemistry}\ }\textbf {\bibinfo {volume} {61}},\
  \bibinfo {pages} {191} (\bibinfo {year} {2010})}\BibitemShut {NoStop}%
\bibitem [{\citenamefont {Scalliet}\ \emph {et~al.}(2022)\citenamefont
  {Scalliet}, \citenamefont {Guiselin},\ and\ \citenamefont
  {Berthier}}]{scalliet2022thirty}%
  \BibitemOpen
  \bibfield  {author} {\bibinfo {author} {\bibfnamefont {C.}~\bibnamefont
  {Scalliet}}, \bibinfo {author} {\bibfnamefont {B.}~\bibnamefont {Guiselin}},
  \ and\ \bibinfo {author} {\bibfnamefont {L.}~\bibnamefont {Berthier}},\
  }\href@noop {} {\bibfield  {journal} {\bibinfo  {journal} {Physical Review
  X}\ }\textbf {\bibinfo {volume} {12}},\ \bibinfo {pages} {041028} (\bibinfo
  {year} {2022})}\BibitemShut {NoStop}%
\bibitem [{\citenamefont {de~Geus}\ \emph {et~al.}(2024)\citenamefont
  {de~Geus}, \citenamefont {Rosso},\ and\ \citenamefont
  {Wyart}}]{de2024dynamical}%
  \BibitemOpen
  \bibfield  {author} {\bibinfo {author} {\bibfnamefont {T.~W.}\ \bibnamefont
  {de~Geus}}, \bibinfo {author} {\bibfnamefont {A.}~\bibnamefont {Rosso}}, \
  and\ \bibinfo {author} {\bibfnamefont {M.}~\bibnamefont {Wyart}},\
  }\href@noop {} {\bibfield  {journal} {\bibinfo  {journal} {arXiv preprint
  arXiv:2401.09830}\ } (\bibinfo {year} {2024})}\BibitemShut {NoStop}%
\bibitem [{\citenamefont {Berthier}\ \emph {et~al.}(2019)\citenamefont
  {Berthier}, \citenamefont {Biroli}, \citenamefont {Bouchaud},\ and\
  \citenamefont {Tarjus}}]{Berthier19}%
  \BibitemOpen
  \bibfield  {author} {\bibinfo {author} {\bibfnamefont {L.}~\bibnamefont
  {Berthier}}, \bibinfo {author} {\bibfnamefont {G.}~\bibnamefont {Biroli}},
  \bibinfo {author} {\bibfnamefont {J.-P.}\ \bibnamefont {Bouchaud}}, \ and\
  \bibinfo {author} {\bibfnamefont {G.}~\bibnamefont {Tarjus}},\ }\href
  {\doibase 10.1063/1.5086509} {\bibfield  {journal} {\bibinfo  {journal} {J.
  Chem. Phys.}\ }\textbf {\bibinfo {volume} {150}},\ \bibinfo {pages} {094501}
  (\bibinfo {year} {2019})}\BibitemShut {NoStop}%
\bibitem [{\citenamefont {Charbonneau}\ \emph {et~al.}(2014)\citenamefont
  {Charbonneau}, \citenamefont {Kurchan}, \citenamefont {Parisi}, \citenamefont
  {Urbani},\ and\ \citenamefont {Zamponi}}]{Charbonneau14a}%
  \BibitemOpen
  \bibfield  {author} {\bibinfo {author} {\bibfnamefont {P.}~\bibnamefont
  {Charbonneau}}, \bibinfo {author} {\bibfnamefont {J.}~\bibnamefont
  {Kurchan}}, \bibinfo {author} {\bibfnamefont {G.}~\bibnamefont {Parisi}},
  \bibinfo {author} {\bibfnamefont {P.}~\bibnamefont {Urbani}}, \ and\ \bibinfo
  {author} {\bibfnamefont {F.}~\bibnamefont {Zamponi}},\ }\href@noop {}
  {\bibfield  {journal} {\bibinfo  {journal} {J. Stat. Mech.}\ }\textbf
  {\bibinfo {volume} {2014}},\ \bibinfo {pages} {10009} (\bibinfo {year}
  {2014})}\BibitemShut {NoStop}%
\bibitem [{\citenamefont {Maimbourg}\ \emph {et~al.}(2016)\citenamefont
  {Maimbourg}, \citenamefont {Kurchan},\ and\ \citenamefont
  {Zamponi}}]{Maimbourg16}%
  \BibitemOpen
  \bibfield  {author} {\bibinfo {author} {\bibfnamefont {T.}~\bibnamefont
  {Maimbourg}}, \bibinfo {author} {\bibfnamefont {J.}~\bibnamefont {Kurchan}},
  \ and\ \bibinfo {author} {\bibfnamefont {F.}~\bibnamefont {Zamponi}},\ }\href
  {\doibase 10.1103/PhysRevLett.116.015902} {\bibfield  {journal} {\bibinfo
  {journal} {Phys. Rev. Lett.}\ }\textbf {\bibinfo {volume} {116}},\ \bibinfo
  {pages} {015902} (\bibinfo {year} {2016})}\BibitemShut {NoStop}%
\bibitem [{\citenamefont {Biroli}\ \emph {et~al.}(2008)\citenamefont {Biroli},
  \citenamefont {Bouchaud}, \citenamefont {Cavagna}, \citenamefont {Grigera},\
  and\ \citenamefont {Verrocchio}}]{biroli2008thermodynamic}%
  \BibitemOpen
  \bibfield  {author} {\bibinfo {author} {\bibfnamefont {G.}~\bibnamefont
  {Biroli}}, \bibinfo {author} {\bibfnamefont {J.-P.}\ \bibnamefont
  {Bouchaud}}, \bibinfo {author} {\bibfnamefont {A.}~\bibnamefont {Cavagna}},
  \bibinfo {author} {\bibfnamefont {T.}~\bibnamefont {Grigera}}, \ and\
  \bibinfo {author} {\bibfnamefont {P.}~\bibnamefont {Verrocchio}},\ }\href
  {\doibase 10.1038/nphys1050} {\bibfield  {journal} {\bibinfo  {journal} {Nat.
  Phys.}\ }\textbf {\bibinfo {volume} {4}},\ \bibinfo {pages} {771} (\bibinfo
  {year} {2008})}\BibitemShut {NoStop}%
\bibitem [{\citenamefont {Kurchan}\ and\ \citenamefont
  {Laloux}(1996)}]{kurchanlaloux}%
  \BibitemOpen
  \bibfield  {author} {\bibinfo {author} {\bibfnamefont {J.}~\bibnamefont
  {Kurchan}}\ and\ \bibinfo {author} {\bibfnamefont {L.}~\bibnamefont
  {Laloux}},\ }\href@noop {} {\bibfield  {journal} {\bibinfo  {journal} {J.
  Phys. A}\ }\textbf {\bibinfo {volume} {29}},\ \bibinfo {pages} {1929}
  (\bibinfo {year} {1996})}\BibitemShut {NoStop}%
\bibitem [{\citenamefont {Biroli}\ \emph {et~al.}(2006)\citenamefont {Biroli},
  \citenamefont {Bouchaud}, \citenamefont {Miyazaki},\ and\ \citenamefont
  {Reichman}}]{Biroli06}%
  \BibitemOpen
  \bibfield  {author} {\bibinfo {author} {\bibfnamefont {G.}~\bibnamefont
  {Biroli}}, \bibinfo {author} {\bibfnamefont {J.-P.}\ \bibnamefont
  {Bouchaud}}, \bibinfo {author} {\bibfnamefont {K.}~\bibnamefont {Miyazaki}},
  \ and\ \bibinfo {author} {\bibfnamefont {D.}~\bibnamefont {Reichman}},\
  }\href {\doibase 10.1103/PhysRevLett.97.195701} {\bibfield  {journal}
  {\bibinfo  {journal} {Phys. Rev. Lett.}\ }\textbf {\bibinfo {volume} {97}},\
  \bibinfo {pages} {195701} (\bibinfo {year} {2006})}\BibitemShut {NoStop}%
\bibitem [{\citenamefont {G{\"o}tze}(2008)}]{gotze2008complex}%
  \BibitemOpen
  \bibfield  {author} {\bibinfo {author} {\bibfnamefont {W.}~\bibnamefont
  {G{\"o}tze}},\ }\href@noop {} {\emph {\bibinfo {title} {Complex dynamics of
  glass-forming liquids: A mode-coupling theory}}},\ Vol.\ \bibinfo {volume}
  {143}\ (\bibinfo  {publisher} {OUP Oxford},\ \bibinfo {year}
  {2008})\BibitemShut {NoStop}%
\bibitem [{\citenamefont {Widmer-Cooper}\ \emph {et~al.}(2004)\citenamefont
  {Widmer-Cooper}, \citenamefont {Harrowell},\ and\ \citenamefont
  {Fynewever}}]{widmer2004reproducible}%
  \BibitemOpen
  \bibfield  {author} {\bibinfo {author} {\bibfnamefont {A.}~\bibnamefont
  {Widmer-Cooper}}, \bibinfo {author} {\bibfnamefont {P.}~\bibnamefont
  {Harrowell}}, \ and\ \bibinfo {author} {\bibfnamefont {H.}~\bibnamefont
  {Fynewever}},\ }\href@noop {} {\bibfield  {journal} {\bibinfo  {journal}
  {Physical review letters}\ }\textbf {\bibinfo {volume} {93}},\ \bibinfo
  {pages} {135701} (\bibinfo {year} {2004})}\BibitemShut {NoStop}%
\bibitem [{\citenamefont {Alkemade}\ \emph {et~al.}(2022)\citenamefont
  {Alkemade}, \citenamefont {Boattini}, \citenamefont {Filion},\ and\
  \citenamefont {Smallenburg}}]{alkemade2022comparing}%
  \BibitemOpen
  \bibfield  {author} {\bibinfo {author} {\bibfnamefont {R.~M.}\ \bibnamefont
  {Alkemade}}, \bibinfo {author} {\bibfnamefont {E.}~\bibnamefont {Boattini}},
  \bibinfo {author} {\bibfnamefont {L.}~\bibnamefont {Filion}}, \ and\ \bibinfo
  {author} {\bibfnamefont {F.}~\bibnamefont {Smallenburg}},\ }\href@noop {}
  {\bibfield  {journal} {\bibinfo  {journal} {The Journal of Chemical Physics}\
  }\textbf {\bibinfo {volume} {156}},\ \bibinfo {pages} {204503} (\bibinfo
  {year} {2022})}\BibitemShut {NoStop}%
\bibitem [{\citenamefont {Schoenholz}\ \emph {et~al.}(2017)\citenamefont
  {Schoenholz}, \citenamefont {Cubuk}, \citenamefont {Kaxiras},\ and\
  \citenamefont {Liu}}]{schoenholz2017relationship}%
  \BibitemOpen
  \bibfield  {author} {\bibinfo {author} {\bibfnamefont {S.~S.}\ \bibnamefont
  {Schoenholz}}, \bibinfo {author} {\bibfnamefont {E.~D.}\ \bibnamefont
  {Cubuk}}, \bibinfo {author} {\bibfnamefont {E.}~\bibnamefont {Kaxiras}}, \
  and\ \bibinfo {author} {\bibfnamefont {A.~J.}\ \bibnamefont {Liu}},\
  }\href@noop {} {\bibfield  {journal} {\bibinfo  {journal} {Proceedings of the
  National Academy of Sciences}\ }\textbf {\bibinfo {volume} {114}},\ \bibinfo
  {pages} {263} (\bibinfo {year} {2017})}\BibitemShut {NoStop}%
\bibitem [{\citenamefont {Toninelli}\ \emph {et~al.}(2006)\citenamefont
  {Toninelli}, \citenamefont {Biroli},\ and\ \citenamefont
  {Fisher}}]{toninelli2006jamming}%
  \BibitemOpen
  \bibfield  {author} {\bibinfo {author} {\bibfnamefont {C.}~\bibnamefont
  {Toninelli}}, \bibinfo {author} {\bibfnamefont {G.}~\bibnamefont {Biroli}}, \
  and\ \bibinfo {author} {\bibfnamefont {D.~S.}\ \bibnamefont {Fisher}},\
  }\href@noop {} {\bibfield  {journal} {\bibinfo  {journal} {Physical review
  letters}\ }\textbf {\bibinfo {volume} {96}},\ \bibinfo {pages} {035702}
  (\bibinfo {year} {2006})}\BibitemShut {NoStop}%
\bibitem [{\citenamefont {Keys}\ \emph {et~al.}(2011)\citenamefont {Keys},
  \citenamefont {Hedges}, \citenamefont {Garrahan}, \citenamefont {Glotzer},\
  and\ \citenamefont {Chandler}}]{keys2011excitations}%
  \BibitemOpen
  \bibfield  {author} {\bibinfo {author} {\bibfnamefont {A.~S.}\ \bibnamefont
  {Keys}}, \bibinfo {author} {\bibfnamefont {L.~O.}\ \bibnamefont {Hedges}},
  \bibinfo {author} {\bibfnamefont {J.~P.}\ \bibnamefont {Garrahan}}, \bibinfo
  {author} {\bibfnamefont {S.~C.}\ \bibnamefont {Glotzer}}, \ and\ \bibinfo
  {author} {\bibfnamefont {D.}~\bibnamefont {Chandler}},\ }\href@noop {}
  {\bibfield  {journal} {\bibinfo  {journal} {Physical Review X}\ }\textbf
  {\bibinfo {volume} {1}},\ \bibinfo {pages} {021013} (\bibinfo {year}
  {2011})}\BibitemShut {NoStop}%
\bibitem [{\citenamefont {Puosi}\ and\ \citenamefont
  {Leporini}(2012)}]{puosi2012communication}%
  \BibitemOpen
  \bibfield  {author} {\bibinfo {author} {\bibfnamefont {F.}~\bibnamefont
  {Puosi}}\ and\ \bibinfo {author} {\bibfnamefont {D.}~\bibnamefont
  {Leporini}},\ }\href@noop {} {\bibfield  {journal} {\bibinfo  {journal} {The
  Journal of chemical physics}\ }\textbf {\bibinfo {volume} {136}},\ \bibinfo
  {pages} {041104} (\bibinfo {year} {2012})}\BibitemShut {NoStop}%
\bibitem [{\citenamefont {Berthier}\ \emph {et~al.}(2012)\citenamefont
  {Berthier}, \citenamefont {Biroli}, \citenamefont {Coslovich}, \citenamefont
  {Kob},\ and\ \citenamefont {Toninelli}}]{berthier2012finite}%
  \BibitemOpen
  \bibfield  {author} {\bibinfo {author} {\bibfnamefont {L.}~\bibnamefont
  {Berthier}}, \bibinfo {author} {\bibfnamefont {G.}~\bibnamefont {Biroli}},
  \bibinfo {author} {\bibfnamefont {D.}~\bibnamefont {Coslovich}}, \bibinfo
  {author} {\bibfnamefont {W.}~\bibnamefont {Kob}}, \ and\ \bibinfo {author}
  {\bibfnamefont {C.}~\bibnamefont {Toninelli}},\ }\href@noop {} {\bibfield
  {journal} {\bibinfo  {journal} {Physical Review E}\ }\textbf {\bibinfo
  {volume} {86}},\ \bibinfo {pages} {031502} (\bibinfo {year}
  {2012})}\BibitemShut {NoStop}%
\end{thebibliography}%
		
		\appendix
		
		\section{Numerical Details}
		\label{det}
		
		We use for system sizes $N_{d=2}=484$ and $N_{d=3}=512$, which allow to perform extensive simulations as  $\phi$, $f_P$ and $f_R$ are varied, while having small finite-size effects for the dynamical range considered \cite{berthier2012finite}. Our Monte Carlo algorithm follows previous choices \cite{ninarello2017models}: it involves displacements and swap moves, where the latter are attempted with a probability of $20\%$. The magnitude $\delta l$ of translation moves is chosen such  that the acceptance ratio is $75\%$.
		
		To achieve rapidly equilibration for any choice of $(\phi,f_R,f_P)$, we first use our fastest Monte Carlo, corresponding to $(\phi,f_R=0,f_P=0)$. Then our algorithm is run with the desired values of $(f_R,f_P)$ for $10^9$ Monte Carlo steps. To check that equilibration was reached, we compare relevant observables (such as correlation functions)
		in the first and second half of the run, and test for consistency.
		
\begin{figure}[h]
			\centering
			\includegraphics[width=0.98\columnwidth]{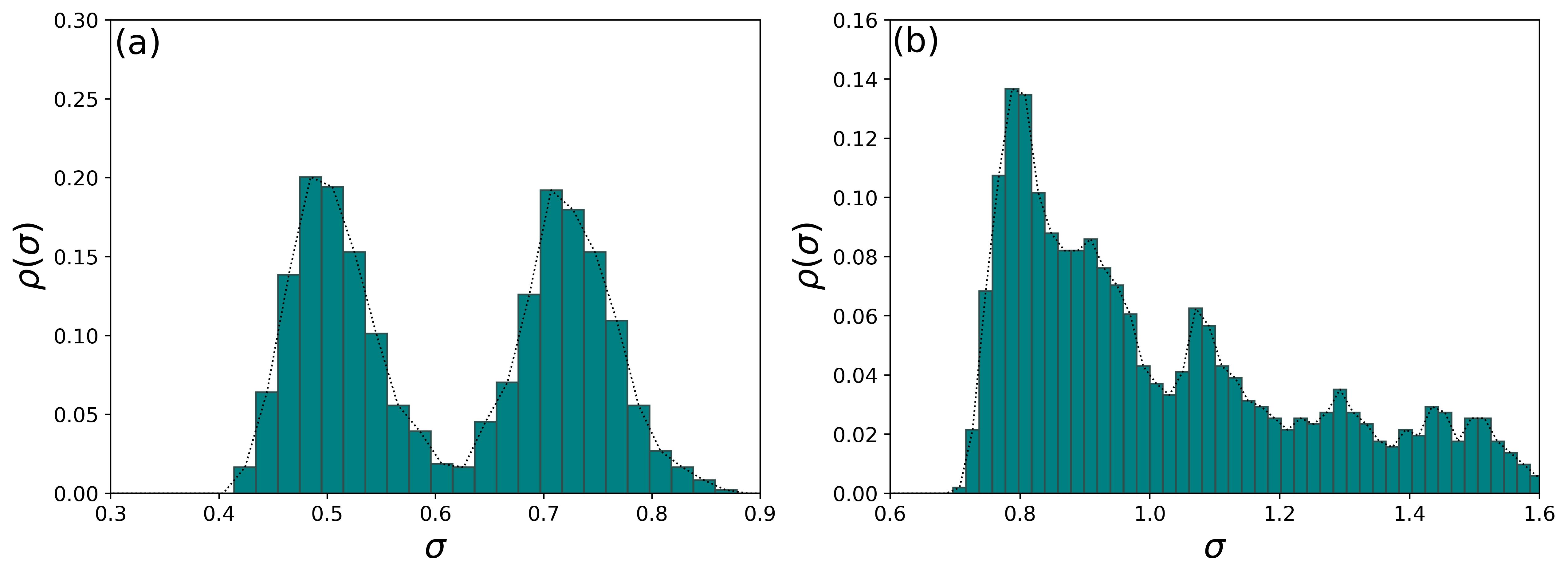}
			\caption{Initial radius distribution for the (a) 2D system and (b) 3D system. 
			}
			\label{hist}
		\end{figure}

		The magnitude of the effect of the SWAP in speeding up the dynamics is affected by the width of the particle radii distribution. The degree of polydispersity is quantified as in \cite{ninarello2017models} by some quantity   $\delta = \sqrt{\langle \sigma^2 \rangle - \langle \sigma \rangle ^2}/\langle \sigma \rangle$, where  $\langle \sigma \rangle$ is the average diameter of the packing which our unit length. In this work we use packings with sizes distribution shown in Fig. \ref{hist}, which have polidispersity $19\%$ and $23\%$ for $d=2$ and $d=3$  respectively.
		
		\section{The effect of the continuous swap}
		\label{appContSwap}
		
		In order to characterize the dynamics of the system for $d=3$ we consider the usual self-scattering correlation function $ F_s(k,t)=\left\langle\frac{1}{N}\sum_i e^{i\mathbf{k}\cdotp[\mathbf{r}_j(t) - \mathbf{r}_j(0)]}\right\rangle$, where $\mathbf{r}_j(t)$ is the position of the particle $j$ at time $t$ and the wave vector $\mathbf{k}$  satisfies $|\mathbf{k}|=2\pi/\langle \sigma \rangle$. For $d=2$, this definition is not well-suited (long-wavelength vibrational modes  bring $ F_s(k,t)$ to zero for large $t$ and $N$, even in a crystal). This problem is fixed as is usually done by considering the relative motion of particle with respect to their neighbors. It can be achieved by introducing the correlation $C(t)=\frac{1}{N} \sum_i  C^i(t)$, with $C^i(t)=\frac{1}{n_i(t)} \sum_{\langle ij \rangle} W ( 1 - \frac{|(\mathbf{r}_j(t) - \mathbf{r}_j(t_0)) - (\mathbf{r}_i(t) - \mathbf{r}_i(t_0))| }{\langle R \rangle})$ where $\langle R \rangle$ is the average radius of the particles, $n_i(t)$  is the number of neighbors of the particle $i$ at time $t$\-- defined as all particles $j$ for which $|\mathbf{r}_j(t)-\mathbf{r}_i(t)|<2.4\langle R \rangle$, an estimate of the first shell of neighbors, and $W(x)$ is the Heaviside function. To extract a relaxation time $\tau_\alpha$,   $F_s(k,t)$ and $C(t)$ are fitted with an stretched exponential function $f(t) = \exp (-(t/\tau_{\alpha})^\beta)$, where $\tau_{\alpha}$ is the relaxation time. The glass packing fraction $\phi_G(f_R,f_P)$ is then defined such that $\tau_{\alpha}=\tau_{\alpha}^*\equiv 10^7$ Monte Carlo steps per particle. The associated stretch exponents  $\beta_G$ are reported in the Appendix as a function of $(f_R,f_P)$. 
		
		{\bf Speed up:}
		For $d=3$, our choice of polydispersity and Monte-Carlo algorithm follows closely previous works \cite{berthier2016equilibrium, ninarello2017models}, where it was shown that swap can speed up the dynamics by 15 decades or more.  Here we observe a giant speed up in our $d=2$ system as well, that continuously builds up as $f_R$ decreases toward the swap case $f_R=0$ starting from the normal dynamics $f_R=1$. Fig.\ref{cont_swap}, {in Appendix~\ref{appContSwap},} shows the relaxation times $\tau_{\alpha}$ {-- extracted from correlation functions as recalled in Appendix~\ref{strExp} --}  as a function of the packing fraction  $\phi$ for different values of $f_R$. It is notable that $\tau_{\alpha}$   depends very significantly on $f_R$, but that this dependence is continuous. Finally, the speed-up of swap can be approximately extrapolated to the case where the normal dynamics would reach experimental time scales (i.e. would increase by 15 decades). The range of the corresponding $\phi_G^{exp}$ can be  estimated by fitting  the curve $\tau_{\alpha}(\phi)$ of the normal dynamics $f_R=1,f_P=0$ with plausible functional forms for $\tau_\alpha(\phi)$.
		The functional forms used to fit $\tau_{\alpha}(\phi)$ and estimate $\phi_G^{exp}$ are the Vogel-Fulcher-Tamman  $\tau_{\alpha}\sim \exp\left( \frac{A}{\phi_{VFT}-\phi}\right)$  or a form with non-singular activation energy $\tau_{\alpha}=\tau'_{\infty}\exp\left( A'(\phi_c-\phi)^2\right)$. 
		In agreement with previous such inferences \cite{berthier2016equilibrium, ninarello2017models}, we obtain that the swap dynamics $f_R=1$ has only slowed-down by 3 to 6 decades at $\phi_G^{exp}$: the speed up conferred by swap is very high.

		\begin{figure}[h]
			\includegraphics[width=0.8\columnwidth]{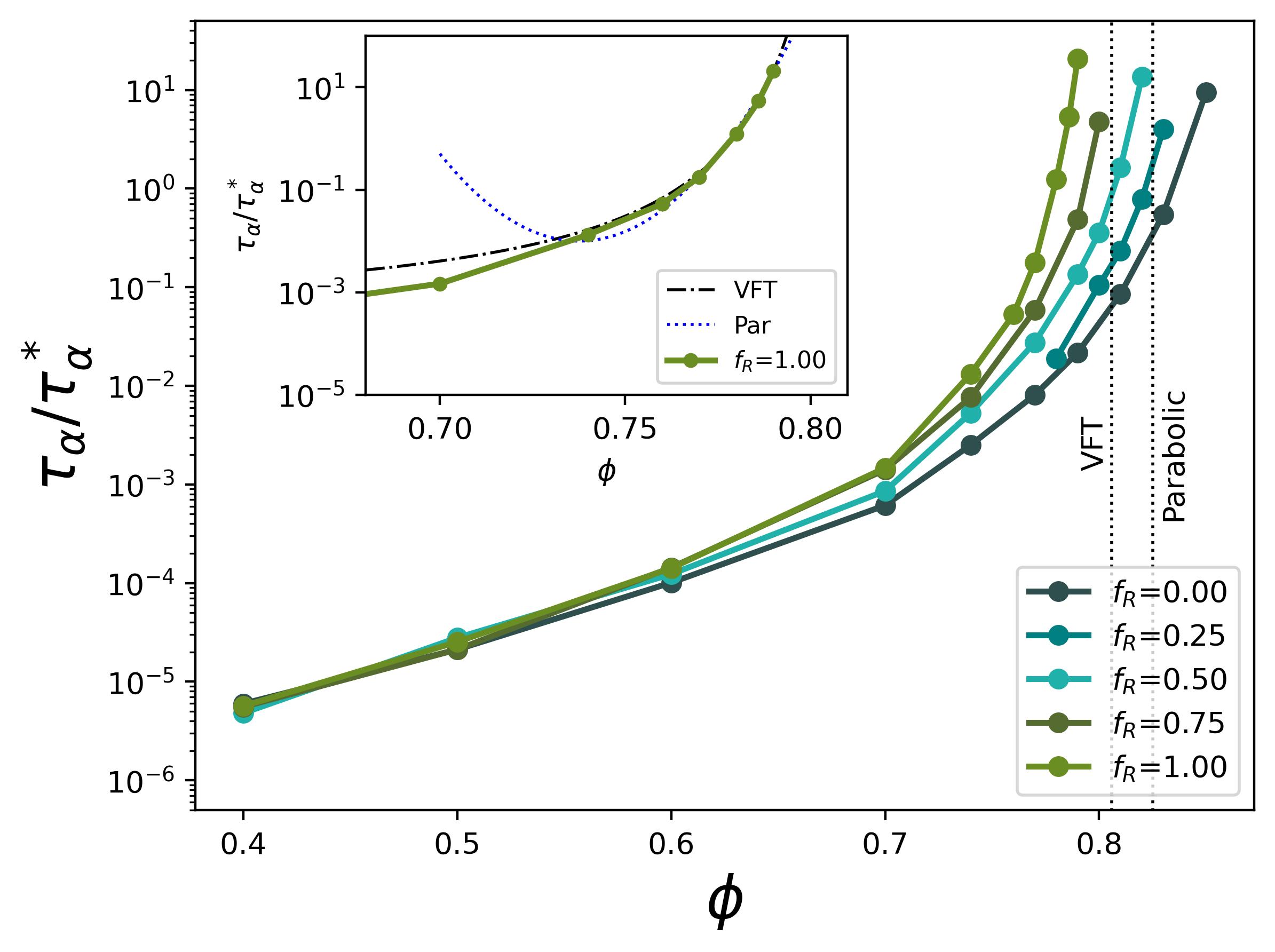}
			\caption{$\tau_{\alpha}/\tau_{\alpha}^*$ as a function of $\phi$ for a two-dimensional system for different values of $f_R$. The dotted lines shows the values of $\phi_{VFT}$ and $\phi_{P}$ obtained by fitting the normal dynamics simulation  ($f_R=1$). These fits are shown the insert panel. }
			\label{cont_swap}
		\end{figure}

		
		
			
			\section{Stretched exponents}
			\label{strExp}
			
\begin{figure}[H]
				\centering
				\includegraphics[width=0.98\columnwidth]{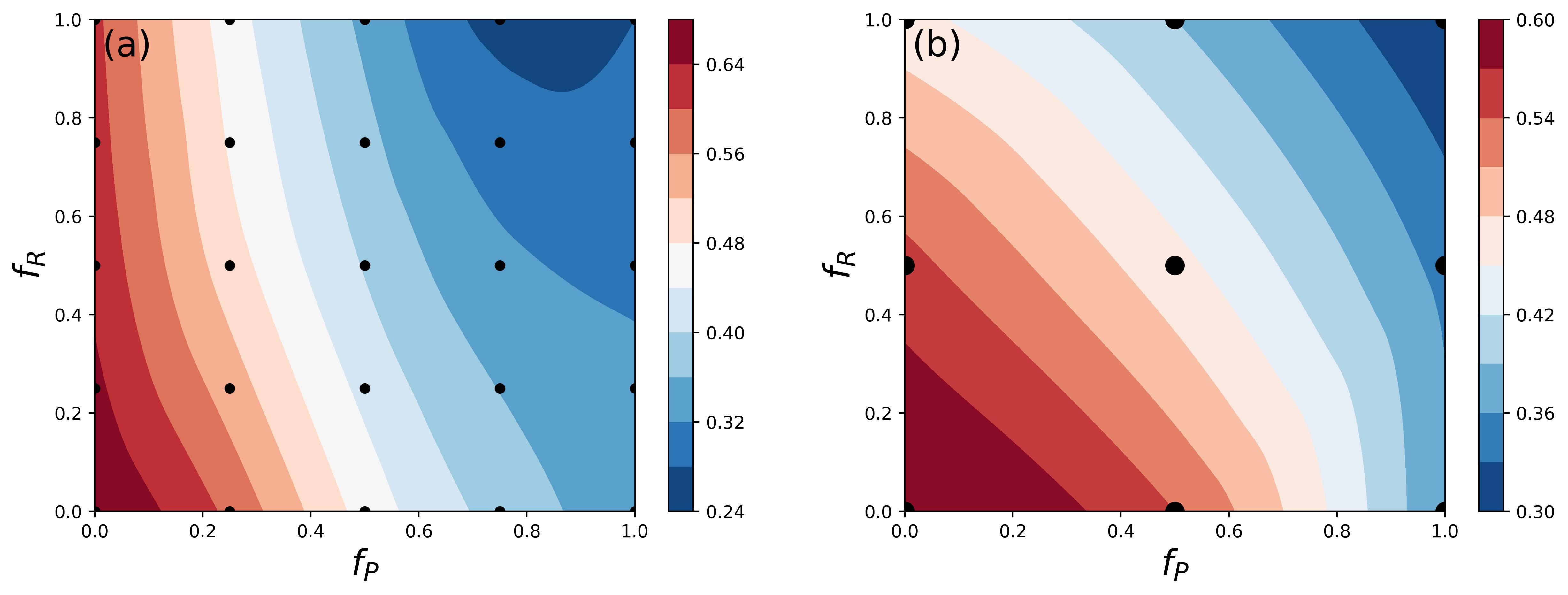}    
				\caption{$\beta_G$ as a function of $f_R$ and $f_P$ for $d=2$ (a) and $d=3$ (b). The color-code represents the value of $\beta_G$. }
				\label{beta}
			\end{figure}

			Figures \ref{beta}-(a,b) show $\beta_G$ in color as a function of $f_R$ and $f_P$ for 2D and 3D system respectively where $\beta_G$ is the value of $\beta$ obtained at $\phi_G$. These values were obtained through the fitting of function $F_s(k,t)$ and $C(t)$ with an stretched exponential function $f(t) = exp (-(t/\tau_{\alpha})^\beta)$, where $\tau_{\alpha}$ is the relaxation time. In both cases $\beta_G$ decreases as the system becomes more restricted. 
			
			\section{Measure of $l_c(t)$ }
			\label{lc}
			
\begin{figure}[H]
				\centering
				\includegraphics[width=0.95\columnwidth]{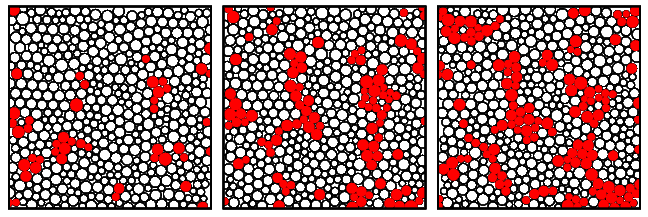} 
				\caption{Spatio-temporal evolution of the structural relaxation in a 2d,  $f_R=0, f_P=0$ and $\phi=0.79$, for which $\tau_{\alpha}=2.6\times10^5$.  Red color indicates relaxed particles. From left to right, the snapshots are taken at times $\tau_{\alpha}/10$ ($l_c\approx 2.7$),  $\tau_{\alpha}/5$ ($l_c\approx 4.5$) and  $\tau_{\alpha}/4$ ($l_c\approx 5.7$). }
				\label{coarsening}
			\end{figure}

			To define a coarsening length $l_c(t)$, we first define relaxed particles.  Particle $i$ is a relaxed particle if $C^i(t) \le 0.5$. Recall that $C^i(t)$ represents the proportion of particles $j$ that remain neighbors of particle $i$ after a time $t$. Then we consider that two relaxed particles  $i$ and $j$ belong to the same cluster if $|\mathbf{r}_j(t)-\mathbf{r}_i(t)|<2.4\langle R \rangle$. 
				An example of a growing length scale is presented in Fig.(\ref{coarsening}) where relaxed particles are shown in red. 
			
			\end{document}